# The 2014 Magnetism Roadmap

Robert L Stamps[1], Stephan Breitkreutz[2], Johan Åkerman[3,4], Andrii V Chumak[5], YoshiChika Otani[6,7], Gerrit E W Bauer[8,9], Jan-Ulrich Thiele[10], Martin Bowen[11], Sara A Majetich[12], Mathias Kläui[13], Ioan Lucian Prejbeanu[14,15,16], Bernard Dieny[14,15,16], Nora M Dempsey[17,18] and Burkard Hillebrands[5]

[1] School of Physics and Astronomy, University of Glasgow, Glasgow, UK, G12 8QQ
[2] Technische Universität München, Lehrstuhl für Technische Elektronik, Arcisstrasse 21, Munich, 80333, Germany
[3] Department of Physics, University of Gothenburg, Fysikgränd, Gothenburg, 41296, Sweden
[4] KTH-Royal Institute of Technology, Materials Physics, Electrum 229, Kista, 164 40, Sweden
[5] Fachbereich Physik, Technische Universität Kaiserslautern, Erwin-Schrödinger-Str. 56, 67663 Kaiserslautern, Germany
[6] Institute for Solid State Physics, The University of Tokyo, Kashiwanoha 5-1-5, Kashiwa, Chiba, 277–8581, Japan
[7] Quantum Nano-Scale Magnetism Team, Center for Emergent Matter Science (CEMS), RIKEN, 2-1 Hirosawa, Wako 351-0198, Japan
[8] Institute for Materials Research and WPI-AIMR, Tohoku University, Aoba-ku, Katahira 2-1-1, Sendai 980–8577, Japan
[9] Kavli Institute of Nanoscience Delft, Delft University of Technology, Lorentzweg 1, 2628 CJ Delft, The Netherlands
[10] Seagate Technology, 47010 Kato Road, Fremont, CA, 94538, USA
[11] Institut de Physique et Chimie des Materiaux de Strasbourg, CNRS-U. de Strasbourg, UMR 7504, 23 rue du Loess, BP 43, F-67034 Strasbourg Cedex 2, France
[12] Department of Physics, Carnegie-Mellon University, 5000 Forbes Avenue, Pittsburgh, PA, 1521, USA
[13] Institute of Physics and Graduate School of Excellence Materials Science in Mainz, Johannes Gutenberg Universität Mainz, Staudinger Weg 7, 55128, Mainz, Germany
[14] Univ. Grenoble Alpes, INAC-SPINTEC, F-38000 Grenoble, France
[15] CEA, INAC-SPINTEC, F-38000 Grenoble, France
[16] CNRS, SPINTEC, F-38000 Grenoble, France
[17] Univ. Grenoble Alpes, Inst NEEL, F-38042 Grenoble, France
[18] CNRS, Inst NEEL, F-38042 Grenoble, France

E-mails: robert.stamps@glasgow.ac.uk, stephan.breitkreutz@tum.de, johan.akerman@physics.gu.se / akerman1@kth.se, chumak@physik.uni-kl.de, yotani@issp.u-tokyo.ac.jp, g.e.w.bauer@tudelft.nl, jan-ulrich.thiele@seagate.com, bowen@unistra.fr, sara@cmu.edu, klaeui@uni-mainz.de, lucian.prejbeanu@cea.fr, bernard.dieny@cea.fr, nora.dempsey@neel.cnrs.fr and hilleb@physik.uni-kl.de

## Abstract

Magnetism is a very fascinating and dynamic field. Especially in the last 30 years it has experienced many major advances in the full range from novel fundamental phenomena to new products. Applications such as hard disk drives and magnetic sensors are part of our daily life, and new applications, such as in non-volatile computer random access memory, are expected to surface shortly. Thus it is timely for describing the current status, and current and future challenges in the form of a Roadmap article. This 2014 Magnetism Roadmap provides a view on several selected, currently very active innovative developments. It consists of 12 sections, each written by an expert in the field and addressing a specific subject, with strong emphasize on future potential.

This Roadmap cannot cover the entire field. We have selected several highly relevant areas without attempting to provide a full review – a future update will have room for more topics. The scope covers mostly nano-magnetic phenomena and applications, where surfaces and interfaces provide additional functionality. New developments in fundamental topics such as interacting nano-elements, novel magnon-based spintronics concepts, spin-orbit torques and spin-caloric phenomena are addressed. New materials, such as organic magnetic materials and permanent magnets are covered. New applications are presented such as nano-magnetic logic, non-local and domain-wall based devices, heat-assisted magnetic recording, magnetic random access memory, and applications in biotechnology.

May the Roadmap serve as a guideline for future emerging research directions in modern magnetism.

# Contents

# Interacting nanoelements and nano-structures

*Robert L Stamps*, University of Glasgow

**Status** –Microelectronics is an example of how new functionalities emerge from engineered combinations of carefully chosen material building blocks. Interfaces govern interactions between the building blocks, and much of the technological innovation and evolution of solid state devices has followed improvements in thin film and heterostructure growth capabilities to sculpt and control interfaces. This, together with improved technologies to create complex microstructured platforms, has enabled reliable and economical integration of metallic, semi-conducting and oxide components into robust devices.

Fabrication technologies are now being extended to the broad class of ferroic materials, including magnetic elements, alloys and compounds. Advances in lithography used for defining and shaping interfaces in these materials have created new directions for materials development. Response to electric and magnetic fields are of particular importance for many applications, and functional ferroics containing magnetic components offer useful properties for electronic, optical and radio frequency applications. Examples of applications for ferroic composites include sensors, transducers, filters, oscillators, and spintronic devices and information storage.[1] A full description of electric field effects observed in magnetic systems is beyond the scope of the present article, and for the remainder we focus on magnetic examples only.

The breadth of new directions and possibilities enabled by controlled patterning is nicely illustrated with schemes for magnetic based logic. An early approach for magnetic logic was based on the control of domain wall movement in special wire geometries and arrays of coupled magnetic nanoelements.[2,3] Also, spin wave based logic gates were envisaged, utilising spin wave interferometry. An example of a one-input NOT gate was presented in [4] and is based upon manipulation of spinwave phase.

An enticing aspect of this technology is the possibility to create alternatives to semiconducting micro-electronics based computing, with potential for highly efficient, ultra-low power operation. Schemes based upon magnetic reversal can, in principle, operate in the theoretical limit of thermodynamic efficiency, limited only by the entropy of information.[5]

The search for new magnetic materials for spin wave applications has created its own field, now called *magnonics*. In this field, lithography is used to create arrays of magnetic elements in which high frequency

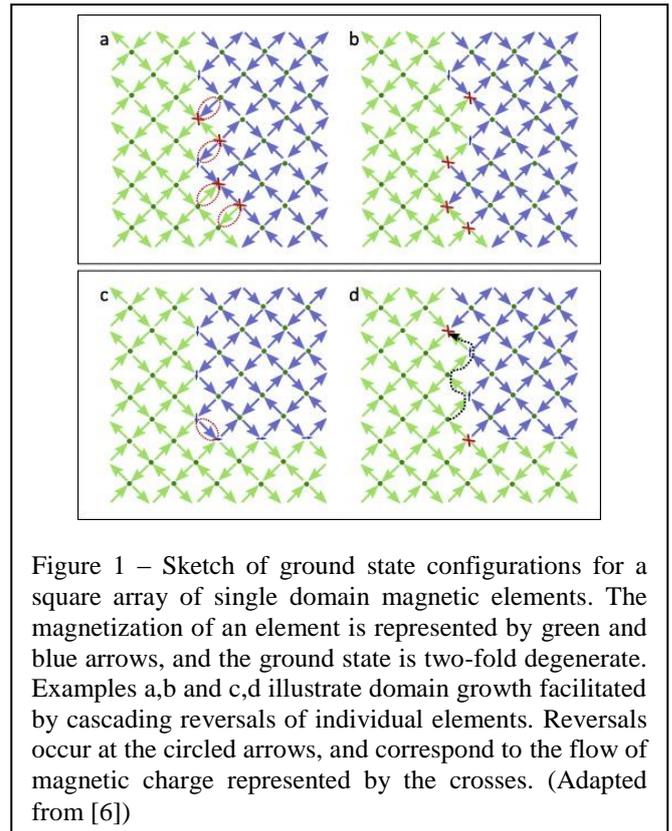

Figure 1 – Sketch of ground state configurations for a square array of single domain magnetic elements. The magnetization of an element is represented by green and blue arrows, and the ground state is two-fold degenerate. Examples a,b and c,d illustrate domain growth facilitated by cascading reversals of individual elements. Reversals occur at the circled arrows, and correspond to the flow of magnetic charge represented by the crosses. (Adapted from [6])

properties are modified by array design. The precise engineering of composite materials is itself a burgeoning field in that nearly every useful ferroic property can be modified and tuned. Structured composites can be designed to create new advanced materials, and the fruit of this technology will be materials engineering of new, useful, artificial ferroic heterostructures.

**Current and Future Challenges** –One approach to strategic design and engineering of materials is to conceive patterned arrays as mesoscopic artificial ferroic materials. As an example, strong dipolar coupling between nano-magnets has been used to create analogues to solid state magnets. A four sublattice antiferromagnet arrangement, as shown in Figure 1, illustrates two degenerate ground states separated by a boundary arrangement.[6] The arrows represent single domain particles that mimic Ising spins as macroscopic moments with a strong uniaxial anisotropy. This system is an example of an artificial magnet, with mesoscopic spin configurations that may have applications in data storage, computation, and microwave signal processing.[7]

The example shown in Figure 1 is a particular realisation of arrangements of magnetic elements that is part of a much larger family of *artificial spin systems* which encompasses several possible arrangements of dipolar coupled nanomagnets. It is possible to modify magnetization properties and processes in these structures to a degree not imaginable in naturally occurring magnetic systems. An essential and far reaching challenge is the fabrication of perfect sub-micron sized magnetic elements and arrays, and development of methods to manipulate, access, and control their magnetic configurations.

Useful ferroic properties may arise on different time and length scales, and these scales can be controlled through composition and design of elements and arrays. In this way, properties of ferroic structures patterned into single domain elements with individual volumes on the order of $10^5$ nm$^3$, can support responses to electric and magnetic fields that occur with characteristic times in the range $10^{-8}$ to $10^{-10}$ s. The key challenges include finding materials and geometries in which surfaces and interfaces can be precisely defined with at least nanometre dimensions. In this way one can also control thermal stability, and construct systems whose functionality derives from operation near the super-paramagnetic limit, as in phase change applications. Additionally, new materials and approaches are required for scalable microwave frequency applications. Such materials require long spin wave propagation path lengths and miniaturization options for use in integrated circuits.

At present controlled fabrication of composites has been realised only for a few material combinations. Next steps include widening the range of materials and exploring more complicated geometries in three dimensions. Such *artificial ferroic systems* can be either entirely magnetic or a mixture of magnetic and ferroelectric constituents, and the structure in these systems is created either during thin film growth or with lithographic methods. Such systems would have many degrees of freedom that are much more complex and tuneable than possible with purely ferromagnetic constituents.

**Advances in Science and Technology to Meet Challenges** –Following the concept of new functionality, an important class of polarisable materials incorporating ferroelectric and magnetic materials could emerge because of dielectric polarisation and strain effects. Here the control of interface structure with complex oxide components is critical in order to introduce desired electric and magnetic field responses. Continued refinements are required, including: tuning of materials properties, use of new materials and improvement of coupling at the ferroelectric and magnetic material interface. Advancement here depends very much on improved control of interfaces, construction of thin film heterostructures incorporating complex oxides, and development of robust lithographic techniques for complex geometries.

An exciting area of development is concerned with the creation of three dimensional mesoscopic ferroic systems. The ability to fabricate arrangements of ferroic elements in complex geometries that go beyond planar arrays is challenging, but within reach. Some examples already exist as concepts for 'racetrack memory' and 'magnetic ratchets' [8,9]. The ratchet concept is based on columnar multilayer elements, and is sketched in Figure 2. In this figure, the formation of solitons in a stack is depicted and indicated by stars. These can be used to represent bits of information and shifted through the stack creating an efficient shift-register in three-dimensional microchips for memory and logic applications.

Further directions impact several different areas. In plasmonics and metamaterials, structures are designed to utilise electromagnetic excitations. For plasmonics, the characteristic length scales of geometrical features should be in the millimetre to micrometre range. A topic to be explored is the incorporation of magnetically polarisable elements in a plasmonic arrays in order to facilitate new and enhanced linear and nonlinear magneto-optical effects [10,11].

Related to this are topics involving charge and spin transport. There are several opportunities for future exploration of transport effects in dipolar coupled magnetic arrays. A particularly exciting direction is the possibility to utilise spin wave excitations in new ways in patterned arrays, as with spin caloritronics in which spin currents can be created by thermal gradients. Since spin waves carry angular momentum, it has been argued that spin-orbit coupling can affect the transport of angular momentum via spin currents across interfaces in some materials. As a consequence, mechanisms originating from spin orbit coupling at interfaces may give rise to spin Seebeck and spin Hall effects, equivalent to their conventional charge transport analogues.[12] This is an exciting and very active area of development at present concerned mainly with thin film structures, but with potentially broad impact on many aspects of magnetization dynamics in patterned nanostructures.

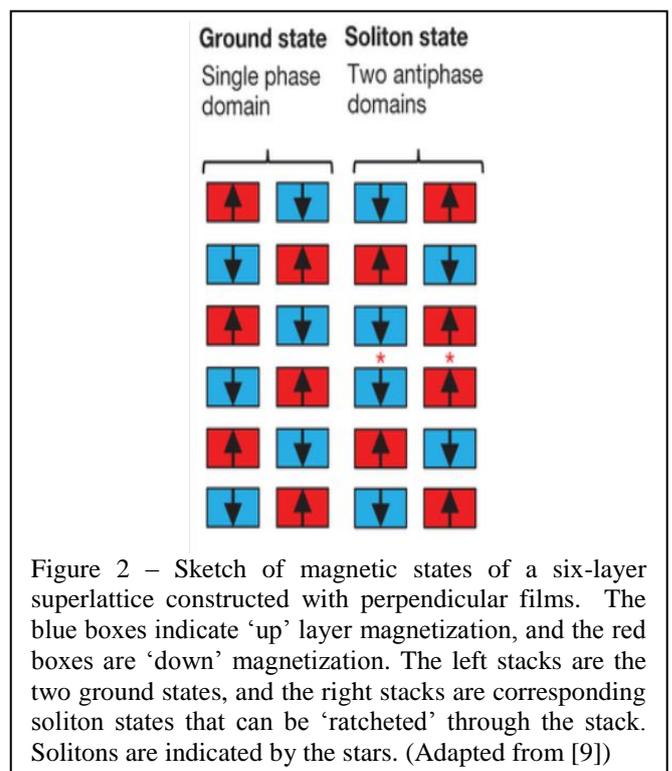

Figure 2 – Sketch of magnetic states of a six-layer superlattice constructed with perpendicular films. The blue boxes indicate 'up' layer magnetization, and the red boxes are 'down' magnetization. The left stacks are the two ground states, and the right stacks are corresponding soliton states that can be 'ratcheted' through the stack. Solitons are indicated by the stars. (Adapted from [9])

**Concluding Remarks** —The ever increasing requirements for information technology and communications include scalability and high density, which necessarily means small size, faster data storage or transfer, higher output signals, low power consumption, low cost, and reproducible control of the magnetic states. New materials that can be easily integrated into existing technologies will be needed to meet these challenges. Artificial ferroic materials based on interacting modular elements provide many new possibilities with which to meet these challenges. Patterned magnetic multiferroics are technologically relevant because of their susceptibility to electric or magnetic fields, thereby allowing these systems to be implemented as devices for data storage, memory and logic, and radio frequency communications. These materials will be created through precision patterning in two and three dimensions, simultaneously advancing fabrication technology into a higher level and creating novel possibilities and opportunities not possible in single phase materials and thin film heterostructures.

# Nanomagnetic logic
*Stephan Breitkreutz,* Technische Universität München

**Status** – Nanomagnetic Logic (NML) is an emerging information processing technology offering nonvolatile logic circuitry to perform boolean and non-boolean operations. Binary information represented by the magnetization state of nanomagnets is processed via field-coupling [1-4]. Nonvolatile majority gate-based logic, low power computing, high density integration, zero leakage and CMOS compatibility are key features of NML. Two different implementations have been developed and established in the research community:

*In-plane NML (iNML)* uses bistable single-domain Py-magnets with shape-dependent in-plane anisotropy arranged in chains and majority gates for logic operation [5]. During clocking, the magnets are forced into a metastable state by a hard-axis clocking field, and, during its removal, switched into their desired state due to fringing field interaction. Meanwhile, the input magnets have to be kept in their desired state to achieve directed signal flow. A 1-bit full adder using slanted-edge magnets has been demonstrated [6].

*Perpendicular NML (pNML)* uses magnets made of Co/Pt multilayer films showing shape-independent perpendicular magnetic anisotropy (PMA) enabling flexible geometries and circuit designs. The PMA is locally tuned by focused ion beam (FIB) irradiation to create artificial nucleation centers (ANCs) for domain walls (DWs). This technique sets the magnet sensitive to only specific neighbors (inputs) and therefore enables directed signal flow integrated in every single magnet [7]. Information is processed stepwise according to the oscillating easy axis clocking field. For switching and correct ordering of the magnets, the clocking field induced DW nucleation at the ANCs is enforced or prevented by the superposing fringing fields of the input magnets. Recently, a threshold logic gate (TLG)-based 1-bit full adder consisting of only 5 magnets has been realized demonstrating the benefits of pNML circuitry in terms of area and power efficiency [8]. A crossing device using signal detouring through additional functional layers is shown in [9].

Both implementations favor Oersted switching and spin transfer torque (STT) devices as electrical I/O structures for integration in hybrid CMOS/NML circuits. For instance, magnetic tunnel junction (MTJ) structures are well studied for devices with in-plane and perpendicular anisotropy. Integrated on-chip coils and current wires generating sub-µs field pulses are used to supply the clocking field [10,11].

**Current and Future Challenges** – For both implementations functional elementary circuits have been experimentally demonstrated, therefore current challenges first of all concern the applicability of NML circuitry. NML has to be engineered to meet the key metrics area, power and speed of potential applications.

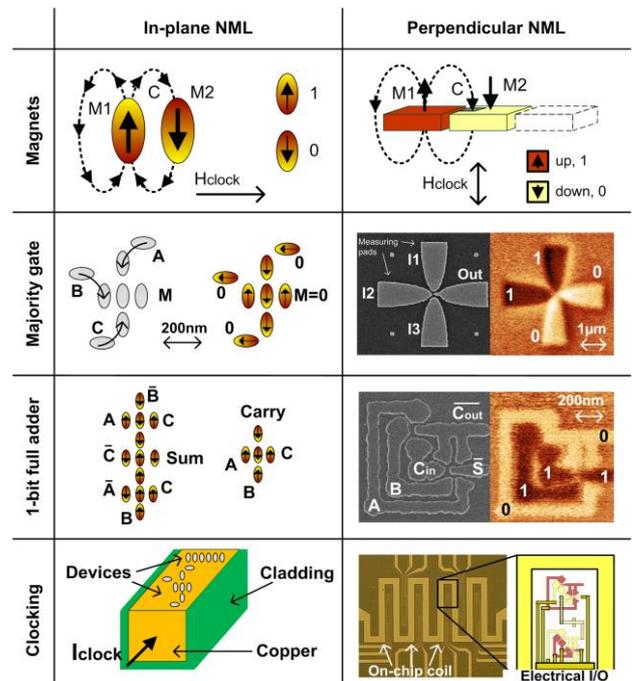

Figure 1 – Comparison of working principle, majority gate, full adder circuit and clocking structure between in-plane and perpendicular NML [5,6,8,10,11].

Hence, research has to focus on material and technological investigations to improve the circuit performance. Additionally, key aspects like signal routing, design and architecture have to be addressed.

*Area*: High density integration is one of the NML key benefits [1]. As the thermal stability of the iNML Py-magnets already reached the $40\,k_b T$ border, there seems to be less potential for further scaling of the magnets themselves [5]. Indeed, design improvements by using the majority decision could further reduce the circuit area [6]. pNML magnets still offer significant scaling potential. Recent publications use magnets with 100 nm width and 30 nm gaps in between [9]. Here, 50 nm wide magnets as demonstrated for magnetic memory devices and 15 nm gaps within logic gates are desirable.

*Power*: The energy dissipation of NML magnets during switching is in the low aJ-range, so there's little potential for further improvement. In fact, NML is one of the most promising low-power technologies [1]. Current research in this area should focus on power reduction of the clocking structures, which contribute for major part to total power losses.

*Speed*: Significant improvements can be gained in the speed of NML circuitry. The (coherent) switching of the iNML magnets occurs below 1 ns, hence GHz clocking frequencies are feasible. By contrast, recent results in pNML indicate working frequencies in the 100 MHz range due to the clocking field generation and the switching process of DW nucleation and propagation [8,10]. Even though data throughput is competitive, speed should be increased to compete with other technologies. For both implementations the required clocking fields also have to allow for GHz operating frequencies.

*Signal routing*: Active control of signal routing is one key challenge for the realization of complex circuitry. Integrated circuits use multiplexers for signal routing and latches to buffer and synchronize data signals. Also in NML circuits, magnetic signals have to be routed, buffered and synchronized between logic units by directly controllable devices. Thus, applicable latches and multiplexer devices have to be developed.

*Architecture*: The nonvolatile character of the magnets makes classical von-Neumann architectures with strict separation of logic and memory redundant. Recent research deals with systolic architectures, non-boolean logic and TLG-based circuitry [4,8]. However, controllable memory access plays an important role also in nonvolatile logic circuitry. Therefore, adequate memory cells have to be implemented.

*Error rate*: One big issue of both NML circuitry is the reliability of magnetic computation that may be impacted due to thermally induced and fabrication variations [8,12]. In addition, the variations of ANCs in pNML circuitry, which are currently manufactured by $Ga^+$ irradiation, are still subject to be reduced. Controlling variations is the key to reduce error rates for highly integrated systems.

## Advances in Science and Technology to Meet Challenges

*Material*: One of the most potential parameters for significant improvements is the magnetic material itself. iNML is still limited to Py, but pNML offers various options of tunable PMA materials as CoNi or CoFeB [13]. They may reduce fabrication dependent variations resulting in lower error rates. Besides that, smaller switching fields due to a lower anisotropy would also strongly reduce the energy consumption of the clocking structures. Higher DW velocities would increase the clocking frequency.

*Technology*: Currently, NML structures are fabricated by E-Beam lithography (both) and FIB irradiation (pNML). However, the manufacturability in industrial plants for mass production is still pending. Significant improvement can be achieved in the irradiation technique of the pNML magnets. Lighter $He^+$ FIB systems or implant technologies with other particles could significantly reduce the variations of the ANCs.

*Clocking*: High permeability low loss claddings for on-chip coils and current wires and enhanced permeability dielectrics (EPD) are currently investigated to reduce the power of integrated NML circuitry [10,11]. Besides common global clocking schemes, innovative clocking concepts using stray fields of propagating DWs to switch nearby magnets seem to be feasible [14], but accurate control of the DWs is required.

*Signal routing*: Here, pNML seems to be the more appropriate candidate due to its switching process. Accurate control of the DW propagation in magnetic nanowires [15] would enable to specifically block the signal flow in magnetic interconnects and therefore enable the routing, buffering and synchronization of magnetic information. By contrast, controlling the signal flow in iNML is more challenging due to the coherent magnet switching. Controllable devices (e.g. integrated in the clocking circuitry [11]) for signal routing and buffering have to be developed.

|  | iNML | pNML |
|---|---|---|
| **Area:** | Decrease of circuit area by majority gate-based design improvements | Reduction of the device & circuit area by scaling of the PMA magnets |
| **Power:** | Reduction of clocking power by device improvements or new clocking concepts (e.g. DW- or voltage-based switching) ||
| **Speed:** | Shorter switching times by material enhancements ||
|  | Higher clocking frequencies by enhanced clocking structures (e.g. use of cladding) or new clocking concepts are required ||
| **Signal routing:** | Concept for iNML circuitry is needed ||
| **Architecture:** | - Systolic computing architectures<br>- 3D integration seems to be feasible ||
| **Error rate:** | Limited decrease by signal refreshment | - Reduction of the ANC variations<br>- Increase coupling |

Table 1 – Current and future challenges for NML circuitry and possible solutions to meet the requirements.

*Architecture*: There are many architectures suitable for NML, but their implementation in NML is still an open question. The usability of systolic and non-boolean computation is currently only confirmed by simulation [4]. A path breaking success is the realization of 3-dimensional magnetic devices [9]. Since field coupling of NML magnets acts in all spatial directions simultaneously, 3D computing devices are feasible.

**Concluding Remarks** – Currently both NML types demonstrated functional basic circuits and electrical integration proving the feasibility of magnetic and hybrid CMOS/NML circuitry. Considering current and future challenges, pNML seems to be the more appropriate candidate: integrated directed signal flow, tunable anisotropy, arbitrary layout, scalability and controllable signal routing are key parameters for integrated circuitry. Computing reliability and speed still have to be increased to compete with other technologies. Controllable devices for signal routing and buffering are required for iNML. Generally, research has to be intensified on suitable materials and fabrication technologies. Furthermore, efforts should be made to reduce the clocking power losses. A fully-integrated NML system including logic, electrical I/O and on-chip clocking would be desirable for the demonstration of the feasibility of integrated circuits.

# Spin torque oscillators
*Johan Åkerman*, University of Gothenburg and KTH Royal Institute of Technology

**Status** – Spin torque oscillators (STOs) [1] are broadband microwave signal generators where a direct current excites and controls a magnetodynamic response via the spin transfer torque (STT) [2,3] effect. STOs typically come in one of two principally different architectures: (i) nano-pillars with a diameter of approximately 100 nm diameter, and (ii) nano-contacts (NC), where the current enters an extended magnetic structure through a constriction. While STOs can be based on single ferromagnetic (FM) layers and multilayers, GMR pseudo spin valves (PSVs) and magnetic tunnel junctions (MTJs), with one free and one fixed layer, dominate. It is advantageous for the free layer to be thin and be made of a material with high spin polarization and low spin wave (SW) damping. For these reasons common free layer materials are sputtered Permalloy ($Ni_{81}Fe_{19}$, here referred to as NiFe) and various compositions of CoFeB, where the former dominates in GMR based STOs and the latter in MTJ based STOs. Owing to their strongly nonlinear magnetodynamics and variety of SW modes, STOs are both of great fundamental interest and lend themselves to potential applications as nanoscale wideband frequency-tunable and rapidly modulated microwave oscillators for telecommunication, vehicle radar, and microwave spectroscopy applications. Additionally, NC-STOs open up for miniaturized integrated magnonic devices and systems where the generation and control of propagating SWs [4] and the recently discovered magnetic droplet soliton [5] and nano-skyrmion [6] may be utilized for additional functionality.

**Current and Future Challenges** - While a wide range of STO device architectures have been fabricated and characterized as a function of drive current and magnetic field, the fundamental properties of the underlying magnetodynamic phenomena and the details of the generated spin wave modes still remain largely hidden. The measured microwave voltage is a strongly reduced mapping of a number of possible highly nonlinear, complex, three-dimensional and time-varying magnetization states. While some spatially resolved information have been possible to extract from scanning micro-Brillouin light scattering [4], and XMCD microscopy [7], the most common approach is still to rely on approximate micromagnetic simulations and compare simulated microwave signals with experiments. The smaller dimensions of nano-pillar STOs and the uniformity of the current density allow for good agreement between simulations and experiments and most fundamental aspects are now well understood. The situation is dramatically different for NC-STOs where much larger simulations must be carried out and where a range of vector and scalar properties are highly non-uniform, such as the current, spin, and heat currents, the spin wave amplitude, and the temperature. The recent advent of powerful GPU accelerated micromagnetic codes have made it possible to reduce the NC-STO simulation times orders of magnitude, resulting in significantly better agreement between simulation and experiments and e.g. some predictive capability of how the non-uniform Oersted field influences the fundamental SW modes underneath the NCs [8]. However, the impact of the non-uniform current density is still largely unexplored, and effective and easy to use fully 3-dimensional simulation tools where both transport and micromagnetics are correctly described are needed. Such simulations will e.g. be critical for modelling single-layer NC-STOs where experiments show that both vortex gyration modes and propagating SW modes can be present and also interact, but likely show different dependence on in- and out-of-plane currents. The recent discoveries of STT generated magnetic droplet solitons [5] and nano-skyrmions [6] also pose additional challenges where additional energy terms such as the Dzialoshinskii-Moria interaction must be properly taken into account in simulations. In addition to improved simulation tools, a better understanding of the spatial and temporal behavior of the generated SW modes will also likely require advances in time-resolved X-ray techniques.

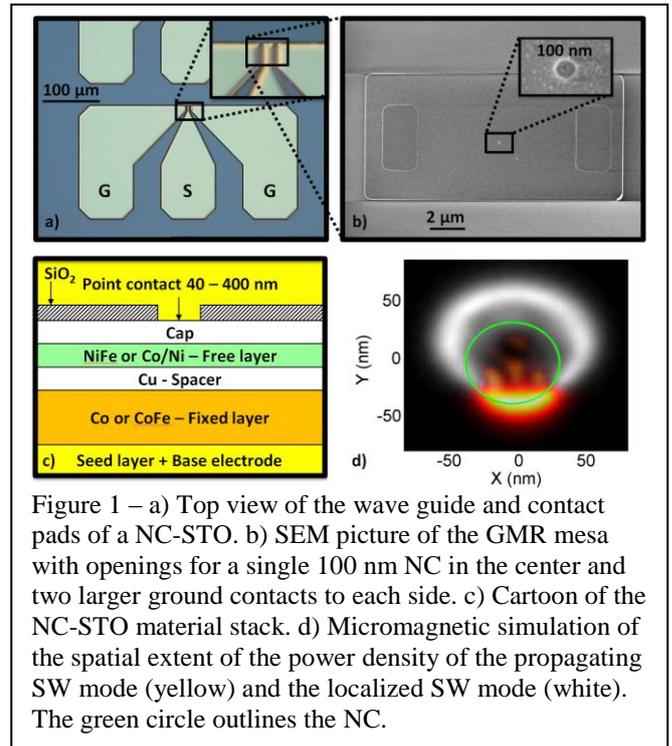

Figure 1 – a) Top view of the wave guide and contact pads of a NC-STO. b) SEM picture of the GMR mesa with openings for a single 100 nm NC in the center and two larger ground contacts to each side. c) Cartoon of the NC-STO material stack. d) Micromagnetic simulation of the spatial extent of the power density of the propagating SW mode (yellow) and the localized SW mode (white). The green circle outlines the NC.

For actual applications, three significant drawbacks still limit the usefulness of STOs: (i) their limited output power, (ii) their high phase noise, and (iii) the common requirement of large magnetic fields, in particular for high-field operation, for high-coherence signal generation, and for magnonic devices relying on propagating SWs. Whereas a common benchmark number in terms of oscillator output power is "0 dBm", which in a 50 Ohm system translates to 1 mW, the highest output power achieved in MTJ based STOs is just above 2 µW [9]. The high phase noise is a result of both the small STO mode volume and the strong intrinsic amplitude-phase coupling which transforms and amplifies all amplitude noise into phase noise. High magnetic fields are typically required to increase the local magnetic energy density determining the frequency of the excited spin waves. The highest quality factors, expressed in line width over operating frequency, are also typically realized in high fields and at out-of-plane field angles tilted some 20-30 degrees away from the perpendicular direction.

**Advances in Science and Technology to Meet Challenges** – The low STO microwave power can be improved upon using a number of approaches. Instead of using GMR in spin valve based STOs, MTJs will have to be used, and their TMR will have to be maximized while simultaneously keeping the resistance-area product (RA) low enough to allow for sufficient current densities to generate coherent auto-oscillations. It will also be necessary to maximize the fraction of the TMR that is actually used by the SW mode. While typical maximum precession angles are of the order of 20 degrees, STOs with perpendicular magnetic anisotropy fixed layers [10] and free layers [11, 12] demonstrate the potential for much larger precession angles. In a similar vein, the recently discovered magnetic droplet soliton [5] can precess at angles of 90 degrees or even higher, i.e. the local spins can precess either in-plane or even against the applied field. Since the STO output power is generated by the variation in resistance during precession, a 90 degree precession angle generates the maximum possible signal. In comparison, a spin precessing at 20 degrees only generates about 12% of that power. Experiments (Fig.2a) indeed demonstrate that the droplet microwave power can be 40 times greater than that of the ordinary precession [5]. If droplets can be generated in MTJ based STOs with a similar increase, the resulting STO power would be about 0.1 mW, i.e. only 10 dB away from the desired 0 dBm level and most likely already sufficient for many applications.

The phase noise can be reduced by increasing the mode volume, e.g. through mutual synchronization of several individual contacts. Recently two important synchronization breakthroughs were reported: *i)* Maehara et al [9] demonstrated mutual synchronization of MTJ based NC-STOs resulting in both record output power above 2 µW and a quality factor of 850, which is considered high for MTJ-STOs; *ii)* Sani et al [10] demonstrated mutual synchronization of three NC-

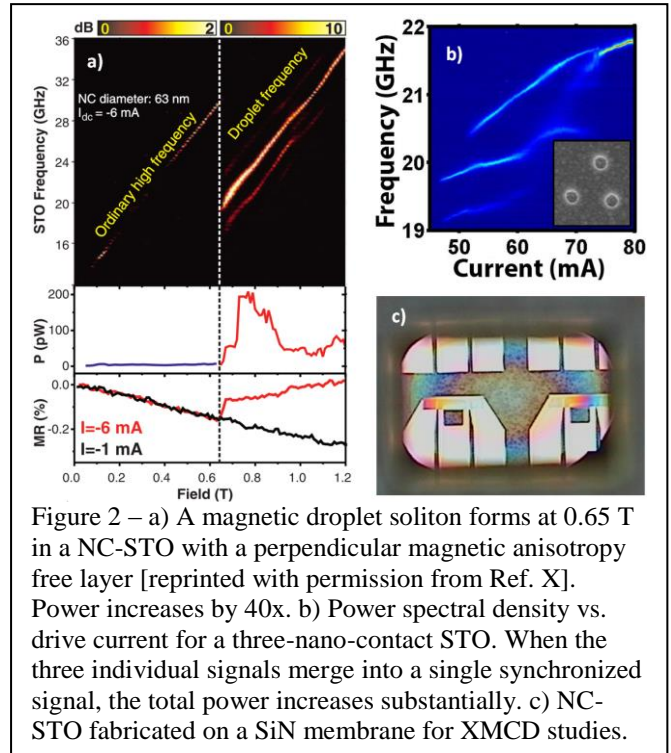

Figure 2 – a) A magnetic droplet soliton forms at 0.65 T in a NC-STO with a perpendicular magnetic anisotropy free layer [reprinted with permission from Ref. X]. Power increases by 40x. b) Power spectral density vs. drive current for a three-nano-contact STO. When the three individual signals merge into a single synchronized signal, the total power increases substantially. c) NC-STO fabricated on a SiN membrane for XMCD studies.

STOs operating above 20 GHz and fabricated using a novel low-cost colloidal lithography approach (Fig.2b). More combined work in these directions will be required to further improve both the output power and phase noise.

On the fundamental side, the still elusive spatial profiles of the various proposed SW modes are currently under intense research by various groups trying to use time-resolved XMCD techniques to acquire direct microscopic information of their spatial extent and wave vector properties. Fig.2c shows an operational NC-STO fabricated on top of a 300 nm SiN membrane currently under investigation using Scanning Transmission X-ray Microscopy and Holographic XMCD with the ambition to directly observe both magnetic droplet solitons and other fundamental SW modes. Alternatively, near-field Brillouin Light Scattering with a resolution of 55 nm [14], and ferromagnetic resonance force microscopy with 100 nm resolution [15], may also help to shed further light on these phenomena.

**Concluding Remarks** – Spin Torque Oscillators remain a vibrant research field with steady progress both in the fundamental understanding and in raising the microwave performance to levels required by applications. The wealth of dynamical phenomena will continue to inspire theoretical, numerical and experimental work, and provide new insights necessary to better tailor these devices and their underlying material properties, both for microwave generators and spin torque driven magnonic devices and circuits.

# Magnon spintronics
*Andrii Chumak*, University of Kaiserslautern

**Status** – Magnons, which are the quanta of the collective excitations of the electrons' spin system – spin waves, was first been predicted by F. Bloch in 1929. Since that time, they are an object of intensive studies oriented towards both fundamental research and application. The wide variety of the linear and non-liner properties as well as the GHz frequency range of magnons used in telecommunication systems, radars, etc., are the main reasons for this interest. The boom in the studies took place after the Yttrium Iron Garnet (YIG) – ferrite with unique small spin-wave losses, was synthesized in 1956. Huge amounts of analog devices for the processing of microwave signals were developed thereafter. Some of them (such as Y-circulators) are still being used today; however, the fast-growing semiconductor technology directed at digital data processing has substituted for the majority.

A renaissance of magnon studies is currently being observed. There are several reasons for this: new technologies allowing for operations with magnons at the nano-scale; a number of discovered physical phenomena such as spin pumping or spin transfer torque (STT) and, finally, a need for an alternative to CMOS technologies due to their fundamental limitations. The term *magnonics* was introduced in 2005 which refers to the transport and processing of data by magnons [1-3]. The usage of magnonics approaches in spintronics (field dealing conventionally with electron-carried spin-currents, see Fig. 1) gave birth to the field of *magnon spintronics*. There are several advantages which magnons offer to spintronics: (1) Magnons allow for the *transport and processing of spin information without the movement of any real particles* such as electrons and, therefore, without the generation of Joule heat. (2) The free path of magnons is usually several orders of magnitude larger compared to the spin diffusion length in metals and thus allow for the *transport of spin information over macroscopic distances*. (3) Finally, the wave nature of spin waves and their abundant non-linear properties provide access to *new and more efficient concepts for data processing*.

**Current and Future Challenges** – Here, studies which are of major importance to the field are listed.
o *New approaches for combining magnonics with spintronics and electronics.* The general concept of magnon spintronics is shown in Fig. 1. The conversion from charge- and spin-currents to magnons and vice versa allows for the combination of different technologies and are of paramount importance. A rather promising type of conversion is based on the combination of spin pumping, STT, and the (inverse) spin Hall effect [4-7]. This approach allows for the conversion by placing a several nanometer-thick non-magnetic metal on top of magnon conduits. Moreover, STT allows for the amplification and generation of magnonic currents [6, 7].
o *New magnetic materials for magnonics.* Magnonics is strongly coupled to Material Science since it requires media with small magnetic damping. The most commonly used materials are Permalloy [3, 8] and YIG [2], but the development of new materials is crucial. For example, it has been shown recently that half-metallic Heusler compounds might be an excellent choice for magnonics not only due to the small damping but also due to the high saturation magnetization and relatively high Curie temperature.
o *Artificial magnetic materials.* Over the last years there have been major achievements in the development and studies of artificial materials with periodic variations of magnetic properties – magnonic crystals. It has been shown that these crystals allow for guiding, filtering, and even short-time storage of magnons on the macro- and micro-scales [2, 3, 8]. Thus, magnonic crystals are very promising for the realization of reconfigurable data processing units. Further achievements in the realization of nano-sized and three-dimensional crystals are expected.
o *Minituarization and increase in the operating frequency.* Decrease in sizes is of highest priority not only for all-metallic magnonics, but also for low-damping YIG devices. Several groups have very recently reported on the growth and usage of high-quality YIG films of nanometer thickness (starting from 4 nm). We have also studied magnons in nanometer-thick YIG waveguides with lateral sizes of a few microns [9]. Decrease in the sizes of structures and spin-wave wavelengths will automatically result in the increase of the spin-wave frequency due to the positive slope of the dispersion of exchange magnons. The minimum wavelength is limited by the lattice constant of a magnetic material, while the maximum magnon frequency reaches the THz range.

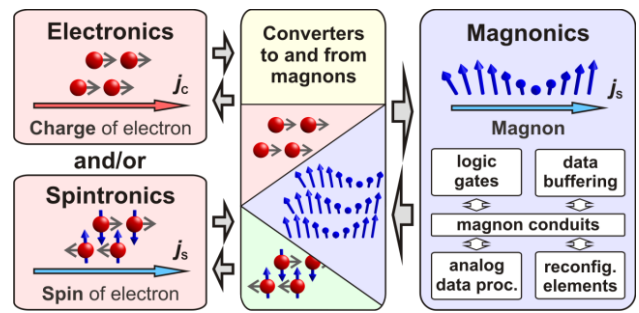

Figure 1 – The concept of magnon spintronics: Information coded into charge- or spin-currents is converted into magnon flows, processed within the magnonic system and converted back. The wave nature of the magnon flow is used for fast processing of complex data inputs.

o *Digital data processing.* Initial steps in the processing of digital information by magnons were taken recently [10]. One of the earliest realizations of a proof-of-principle XNOR logic gate is shown in Fig. 2. A DC current is used to manipulate spin waves and wave interference to perform logic operations. Another promising approach of the coding of information into the spin-wave's phase rather than in its amplitude was proposed by A. Khitun [11]. Furthermore, we have demonstrated that the magnons' natural non-linearity can be used in order to manipulate one magnon with another. A magnon transistor for all-magnon data processing on a single magnetic chip was shown. However, the combination of several logic elements into one magnonic circuit still needs to be realized.

o *New data processing concepts based on linear and non-linear magnon properties.* One of the greatest advantages of magnonics is the possibility to utilize the wave properties at the nanoscale, while most waves of other nature (such as X-rays) are practically inapplicable. This provides access to new data processing concepts in which operations with clusters of information rather than with single bits are used. For example, the linear time reversal of a complex wave packet by a single operation was shown [12]. Fast dynamic magnetic media and nonlinear spin-wave effects suggest huge opportunities for future studies.

o *Magnon manipulation and generation by electric fields.* CMOS technology has shown that the manipulation of information by electric fields consumes minimum power. Besides in the (multi)ferroic-like approaches used in the manipulation of magnons (see for example studies of B. Kalinikos [13]), there is a special interest in the manipulation of magnetic anisotropy in ultrathin films. The group of Y. Suzuki has recently demonstrated low power ferromagnetic resonance excitation by an electric field [14]. Other promising approaches are based on Dzyaloshinskii-Moria interactions but are in the initial stages of development.

o *Data buffering.* Magnons are dynamic objects which require external energy to compensate their damping and are not the best choice for long-term storage of information. Nevertheless, for data processing short term buffering is also required (see Fig. 1). It was shown that the buffering of information carried by magnons is possible for several microseconds using slow magnon modes [2].

o *Magnon caloritronic effects.* In conventional electronics the parasitic heat is ultimately lost. Magnonic systems are also dissipative, but the magnon-phonon interactions are much more sophisticated and include different mechanisms. The very new field of spin caloritronics searches for ways to control or even utilize the parasitic heat [15].

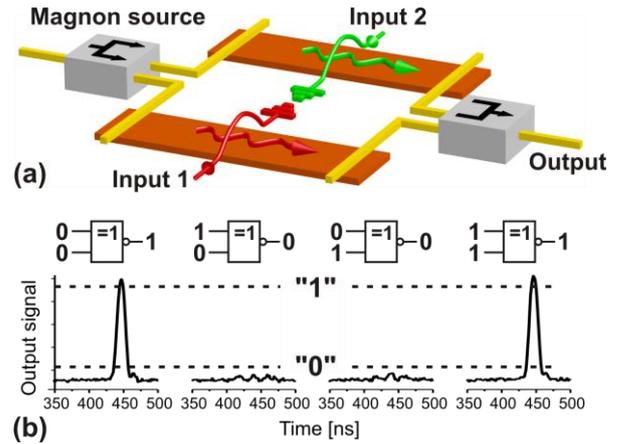

Figure 2 – Spin-wave XNOR gate. The gate is based on a Mach-Zehnder interferometer with arms implemented as spin-wave waveguides [10]. The currents $I_1$ and $I_2$ represent the logical inputs: no current corresponds to logic "0", current $I_\pi$ resulting in a $\pi$-phase shift of spin wave corresponds to "1". The output is given by the amplitude of spin-wave pulse after interference. (b) Measured gate output signals for input signals shown in the diagrams.

o *Quantum effects and Bose-Einstein condensation (BEC) of magnons.* Magnons are bosons with a spin equal to one. It has been shown that magnons can form BEC at room-temperature if the density of the injected magnons is high enough. Currently, there are first attempts to use BEC for the transport and processing of spin information in a form of magnon super-currents. Other operations which are carried by a single magnon and the related quantum effects are also of great importance but so far remain beyond realization.

**Advances in Science and Technology to Meet Challenges** - In general, most of the ingredients required for the successful realization of the intended aims facing magnon spintronics have already been developed: The knowledge of the physics concerning magnons is very comprehensive, modern technology allows for the fabrication of magnonic structures of sub-ten nanometers sizes (for example using a Helium ion microscope). There are also existing techniques which allow for operations in the THz frequency range. The main hurdle in the field of magnonics probably remains the spin-wave relaxation, which has to be decreased utilizing new materials or compensated using new energy-efficient means.

**Concluding Remarks** – Although magnon spintronics is in its initial stages of development there is already an indication for much potential for the development of a particle-less technology in which information is carried and processed by magnons rather than by electrons. Nanoscale, THz frequencies, a rich physics allowing for the development of reconfigurable logic elements and of new types of processors are just some of the many advantages which are proposed by magnons.

# Non-local based devices e.g. spin-orbit torque / spin-orbit based devices

*YoshiChika Otani*, ISSP University of Tokyo & RIKEN – CEMS (350 words max)

**Status** – Non-local based devices can be classified into 4 types, 1) lateral spin valves, 2) lateral switching devices, 3) static spin Hall (SH) devices, and 4) dynamic SH devices. They are expected to become key components in low power memory applications.

The first devices were lateral spin valves (figure 1 (a)) that comprised spin injector ($F_i$) and detector ($F_d$), both made of ferromagnetic nano-wires, bridged by a spin reservoir non-magnetic (N) nano-wire. The scheme is called "non-local" because there is no net flow of charge in N while the devices are in the operation. The magnitude of spin accumulation can be determined by measuring the voltage (V) in lateral spin valves (figure 1 (a)).

Lateral spin valves were modified (figure 1 (b)) to inject pure spin currents into conductive magnetic nano pillars. There spin currents switched magnetization via the spin transfer torque [1]. This offered a new paradigm for manipulating the magnetic moment without applying any magnetic field. Stimulated by the experiments of pure spin current induced magnetization switching (figure 1 (b)), a variety of devices have been proposed, e.g. a three terminal device [2], an all-spin logic device [3] etc.

Separately, static SH devices were developed (figure 1 (c)) to measure charge-spin reciprocal conversion phenomena via spin-orbit interaction (SOI), i.e. direct spin Hall (SH) and its inverse effects. Non-local SH devices have subsequently been employed for systematic exploration of materials which resulted in the discovery of Cu-Bi dilute alloys exhibiting giant SH effects [4].

Recently, dynamic spin pumping by a micro wave resonant cavity was applied to inject spin currents in non-local SH devices. In these so-called dynamic SH devices, alternating spin currents were injected into a spin-orbit material (SOM) via a non-magnetic material as in figure 1 (d) [5].

Note that SH effects have also been measured differently by spin pumping in ferromagnetic/non-magnetic (F/N) bilayers [6]. Furthermore a novel technique called spin torque ferromagnetic resonance was shown to switch magnetization of the ferromagnetic layer or even to oscillate magnetization in a nano-pillar shaped ferromagnet through spin torque mediated by SH effect [7].

**Current and Future Challenges** - The direct and inverse SH effects are now recognized as effective means to exert spin torque on the ferromagnet in the F/N bilayers as well as to detect the spin currents generated by the spin injection or pumping (figure 1 (c, d)). For designing SO based devices, important material parameters are thus the SH angle, i.e. the spin-charge conversion yield that determines the magnitude of the spin torque and the spin diffusion length that limits the size of the devices. Both reported values are however significantly different between static and dynamic spin injection methods. Particularly the spin diffusion length obtained from the dynamic method is an order of magnitude smaller than that from the static one. The discrepancies seem attributable to the type of junction, F/SOM or N/SOM. In the case of F/SOM, the SOM, e.g. Pt is known to have an induced magnetic moment in the vicinity of the junction when in contact with F whereas no moment is induced in N for N/SOM. In the case of some SOMs, the influence of the induced moment on the spin diffusion length needs to be clarified.

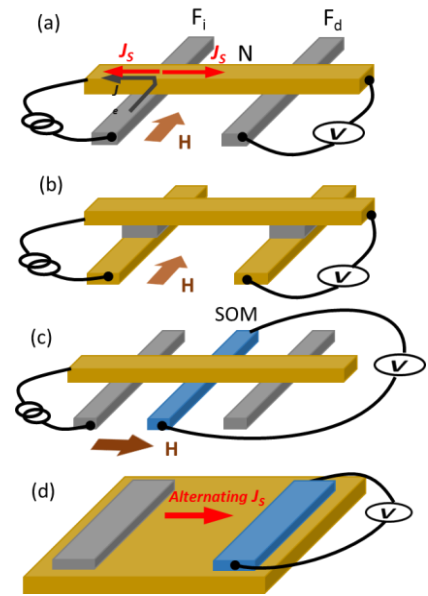

Figure 1. –representative non-local based devices. (a) A lateral spin valve where $F_{i/d}$ and N denote ferromagnetic injector/detector and non-magnetic wires of typically about 100 nm in width, 20~50 nm in thickness, and a few μm in length. (b) A lateral switching device which has a ferromagnetic nano-dots at the junctions for pure spin current induced magnetization switching. (c) A non-local spin absorption device to measure spin charge conversion phenomena for SOMs. (d) A lateral device similar to (c) used for the spin pumping scheme for spin injection. The device is located inside a microwave cavity for operation.

Another important issue is about materials in SH devices (figure 1 (c, d)). So far the SH effects based phenomena have mostly been demonstrated in single 4d and 5d transition metal elements such as Pd, Ta, W and Pt, but not in alloys. There is an experimental demonstration of Cu based dilute alloys [4]. For single elements, tuning the magnitude of the SH angle may only be achieved by changing resistivity because of the intrinsic origin of the SH angle which scales with the resistivity. In contrast, the SH angle for dilute alloys can be tuned in a controlled manner by selecting the combination of impurity and matrix elements.

Apart from SH effects, a new spin charge conversion associated with SOI such as Rashba effect was recently found to take place in Bi/Ag bilayer systems [8]. This implies that interface engineering for manipulating spin-charge conversion is an intriguing and challenging issue for further advancement of conversion efficiency in SH devices. Thorough understanding of the interface contribution may bring about a novel way to exploit spin currents or spin accumulation for controllable physical behaviors other than spin torque induced switching or oscillation.

**Advances in Science and Technology to Meet Challenges** - Besides technical issues concerning the SH angle and the spin diffusion length, the prime task for further advancements in non-local (or SO) based devices may be to explore new mechanisms to manipulate physical phenomena such as the magnetic order of localized spins and the spin life time of conduction electrons by using spin currents or spin accumulation. Two possible new approaches, one using Rashba SOI and the other using non-local spin injection, are described below.

First, recent experimental and theoretical studies on Rashba SOI have been carried out in ferromagnetic thin films [10]. It appeared that the spin accumulation was induced via the interface Rashba effect and that the spin accumulation couples through exchange interaction with localized moments in an adjacent ferromagnetic layer. This is equivalent to the situation where an effective field $B_R$ is applied on the ferromagnetic layer. This effective field amounts to about a few hundreds Oersted which may be useful to induce metamagnetic phase transition, e.g. from antiferromagnetic to ferromagnetic states. A tri-layered structure in figure 2 (a) consisting of a very thin metamagnetic thin layer sandwiched by metal oxide and SOM layers may work as a test device to check the idea.

Second, SH effects could be enhanced in superconducting materials with high SOI such as Nb. Recent experiments revealed that non-local pure spin current injection into a superconducting Nb wire is possible by using a device structure similar to figure 2 (b) [9]. Thereby the intrinsic spin life time in the superconducting Nb was evaluated from the change in spin accumulation and found to become more than 4 times greater than that of the normal state. This enhancement of the spin life time in the superconducting state may bring about a new route for enhancing non-linearly SH effects in non-local based devices.

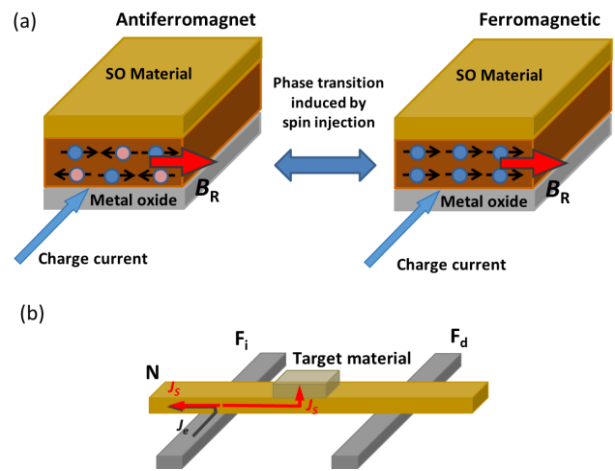

Figure 2 – (a) Schematic representation of magnetic phase transition induced by Rashba effective field $B_R$. (b) Non-local spin injection device structure in which pure spin currents are preferably absorbed into a target material with high SOI.

**Concluding Remarks** – Understanding SOI will pave the way for developing various types of spin-charge conversion devices and will provide seeds for new designs of non-local spin injection devices based on SO torques. Examples include three-terminal devices using ST-FMR for memory applications and spin pumping devices combined with coplanar wave guide for SH effect measurements.

Further progress in the science of non-local based devices is envisaged once new manipulation schemes are established. Two possible approaches for this are given in this article. One is inducing magnetic phase transitions by using Rashba SO effective field, and the other is non-linear enhancement of SH effects in superconducting materials with high SOI.

# Spin caloritronics
*Gerrit E.W. Bauer*, Tohoku University and TU Delft

**Status.** Spin caloritronics is the field devoted to studying coupled heat, charge, and spin currents in magnetic nanostructures. In metallic devices, classical thermoelectric parameters such as the Seebeck and Peltier coefficients develop spin dependence. In magnetic insulators, on the other hand, the physics is quite different; the so-called spin Seebeck effect can be explained as a collective thermomagnetic phenomenon that is caused by spin wave excitations of the ferromagnetic order that pump spin currents into electric contacts. The progress in the field up and including 2011 has been reviewed in Ref. 1, while this note focusses on the level of our understanding and the perceived challenges of spin caloritronics at the end of 2013.

In the past two years great strides have been made to predict, understand, and observe a variety of phenomena. The discovery of the spin Seebeck effect in a non-magnetic semiconductor induced by an applied magnetic field broadened its general appeal [2]. Another highlight is the experimental detection of a spin-heat accumulation. While there is ample evidence for long-lived spin accumulations in which the two spin species have different electrochemical potentials, different temperatures for both spins has been a purely theoretical concept that could only recently be confirmed experimentally in terms of a spin heat-valve effect [3]. A spin-based directional "heat conveyor belt" operated by chiral surface spin waves has been demonstrates on Yttrium Iron Garnet (YIG) [4], a ferromagnetic insulator that is now arguably the most important material for spin caloritronics physics and applications. The spin Peltier effect was expected to exist from reciprocity considerations and has indeed been measured very recently [5] by demonstrating that a current through a Pt contact layer generates an energy current into a YIG substrate, which was found to become warmer or cooler depending on the current direction in Pt.

**Current and Future Challenges.** In spite of the progress reported, there are still many white areas on the spin caloritronics map. In the following I list the ones that in my opinion are most urgent.

- While there is little doubt that heat currents generate spin transfer torques, convincing and reproducible experimental evidence has been hard to come by [6]. Reasons are the complications due to parasitic effects caused by temperature gradients that make a non-equivocal identification of spin transfer difficult. Also the electric excitation of YIG turns out to be recalcitrant. Elucidation of the spin transfer torques delivered to metallic or insulating ferromagnets by heat currents is a very important challenge at present. Clever experimental detective work that separates different thermoelectric, thermomagnetic, and spin caloritronic effects is called for.

- Metal and insulator based spintronics is so attractive for applications because effects are robust with respect to elevated temperatures. Practically all experiments to date have therefore been carried out at ambient conditions. Nevertheless, there might be new physics and specialized applications in the low temperature and high-temperature regimes. At low temperatures thermoelectric effects become small, but tend to give new information, e.g., about many-body effects. New spin caloritronic effects might become observables at low temperatures and quantum structures, also in the presence of superconducting order. On the other hand, applications in the automobile industry require robustness against high temperatures. Nevertheless, both high and low temperature regimes remain virtually unexplored and a challenge for the (near) future.

- The spin Seebeck and spin Peltier effects refers to the spin-heat coupling in bilayers of normal metals and magnetic insulators. In metallic nanostructures spin-dependent thermoelectrics has been invoked to understand the thermopower and cooling by spin-dependent particle currents. However, in metals both effects should coexist and might interact, but a quantitative comparison of the relative importance of single particle thermoelectric and collective effects is still lacking and should be carried out.

- Many recent experiments focus on bilayers with Pt since by the inverse spin Hall effect since is such a convenient spin current detector. However, Pt is a rather extraordinary material and its use has generated some controversies about the microscopic mechanism of different experiments including the longitudinal spin Seebeck effect [7]. Since Pt has a high paramagnetic susceptibility and large spin orbit interaction strength, it is prone to exchange and spin-orbit interaction-induced proximity effects at the interface to ferromagnets. In order to clarify such issues, spin Seebeck and spin pumping experiments should be carried out on samples containing simple metals that are neither spin-polarizable nor relativistic, such as Cu or Al. Spin currents/accumulations can then be detected by ferromagnetic contacts as depicted in Figure 1.

- Another challenge is the physics of ultrafast and highly excited ferromagnets and its heterostructures and its relation to spin caloritronics [8,9]. Also here collective effects due to magnons and particle-based

- spin and heat currents compete. Elucidation of these effects is not trivial because many concepts of near to equilibrium, linear response (Ohm's Law) such as Onsager reciprocity and spin-flip diffusion might become inaccurate or completely inapplicable.

- Magnons are bosons, and therefore amenable to macroscopic condensation. While magnon Bose condensates have been generated by intense microwaves in the past, there is no experimental evidence yet about the expected exotic transport properties such as superfluid spin transport. Thermally facilitated magnon injection from normal metal contact might help to create magnon condensates and study its transport properties [10].

- The spin caloritronics of magnetic superlattices, antiferromagnets, and the role of the staggered order ferrimagnets is another topic of interest.

- While the anomalous and planar Nernst effects have turned out to be ubiquitous and sometimes even a nuisance [11], the complete family of spin caloritronic Hall responses has not yet been observed. The spin Nernst (for a first principles theory see [12]) as well as spin and anomalous Ettingshausen effects have to the best of my knowledge not yet been experimentally identified.

**Advances in Science and Technology to Meet Challenges.** Experiments can help to solve various issues by better control and characterization of interfaces and novel material combinations. On the theory side we need improved band structure calculations for complex magnetic oxides such as YIG and its interfaces with metals, preferably fully including the spin-orbit interaction.

Very important are the efforts to explore the application potential of spin caloritronic principles. By converting the spin Seebeck spin currents the inverse spin Hall effect generates electric power that scales linearly with the device area. The Japanese electronics maker NEC therefore pursues application of the spin Seebeck effect for large area heat scavenging to be used, e.g., in smart textiles [13]. On the other hand, the efficiency of spin caloritronic devices based on the inverse spin Hall effect is small and does not scale in the small sample limit. A spin Seebeck thermoelectric generator in which the thermally induced spin current is detected by ferromagnetic metal contacts generates voltages that do not deteriorate in small structures and may lead to significantly higher thermoelectric figures of merit [14]. The IBM collaboration has modelled a potentially superior spin transfer torque random access memory concept, in which the magnetization reversal is assisted by the spin Seebeck effect [15].

*Concluding Remarks.* Basically all instruments are available to make further progress in spin caloritronics on various fronts; the field is still new and there are many low-hanging fruits to be plucked. This gives ground to optimism that the fast paced progress in recent years will be continued for the time being.

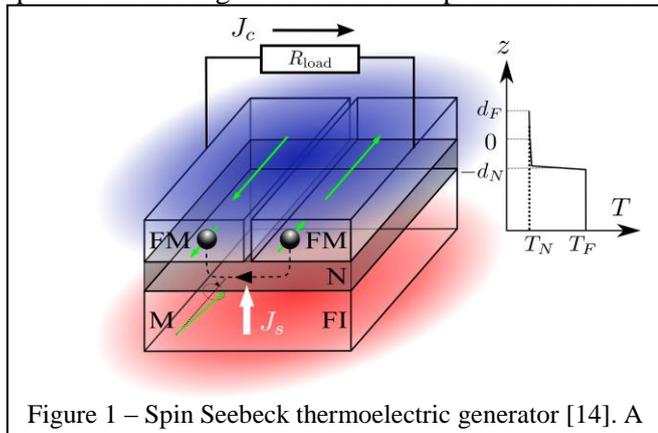

Figure 1 – Spin Seebeck thermoelectric generator [14]. A spin current is generated by the temperature difference between a ferromagnetic insulator FI and normal metal N with weak spin dissipation. The spin accumulation in N is detected by metallic ferromagnetic contacts (FM) with antiparallel magnetizations collinear to that of FI. The spin Seebeck voltage can carry out useful work by generating a charge current through a load resistance.

# Heat Assisted Magnetic Recording (HAMR)
*Jan-Ulrich Thiele*, Seagate Technology

**Status –** The principles of magnetic data storage were invented in the late 19th century, and the venerable magnetic hard disk drive (HDD) has been around for over 50 years, but the insatiable demand for digital data storage fueled by the computer and internet revolutions over the past decades is still driving demand for ever increasing areal density (AD), larger capacities and faster performance of HDDs. The first HDD product, IBM's refrigerator-size 305 RAMAC introduced in 1956 offered a total storage capacity of 4.4 megabytes distributed over 50 double-sided, two-foot-diameter disks at an AD of 2kb/in$^2$ and a price of $10M per gigabyte. By comparison, a modern 2.5" drive for notebook PCs is about the size of a deck of cards, has a capacity of 1 terabyte, stores data at 750 Gb/in$^2$ at a price of less than $0.10 per gigabyte. Despite a number of changes in technology the underlying principle of this $10^8$ increase in AD has been "scaling" of the recording system, i.e., shrinking all critical dimensions of the magnetic write poles, the magnetic read sensor, the magnetic grains making up the recording layer in the media and the mechanical spacing between head and media. Only in the last 15 years have the physical limitations of this trajectory become evident. They have been described as the "trilemma" of magnetic recording, i.e., the balance between three competing requirements: (1) the ability to write data onto the magnetic disk by means of a magnetic field, (2) the ability to retrieve that data with high confidence and speed through a magnetic read sensor, and (3) the retention of the data for the desired life of the HDD. The latter is limited by the so-called super-paramagnetic effect, i.e., the thermal instability of the magnetization direction of an individual magnetic grain in the media when the grain volume is reduced to the point where the magnetic energy per grain, $K_U V$ ($K_U$: magneto-crystalline anisotropy, V: grain volume), becomes comparable with thermal energies, $k_B T$. In media using CoCrPt-based alloys this effect first became noticeable at a magnetic grain diameter of just under 10 nm and an AD of 35 Gb/in$^2$, but progress in tightening distributions of grain sizes and magnetic properties, and higher magnetic fields and field gradients achieved by new write pole geometries used in current state-of-the-art Perpendicular Magnetic Recording (PMR) systems have allowed a 20x increase in AD without further reduction in grain size since then [1].

**Current and Future Challenges** – While magnetic materials with higher anisotropy and therefore larger stability against thermal fluctuations exist, their use has been limited by the maximum magnetic field available from the magnetic materials used for the write poles at ever shrinking pole geometries. Scaling of the recording system has also been slowed down by the inability to further reduce the head-to-media spacing, i.e., the thicknesses of the protective layers on the head and media and the fly height of the aero-dynamic surface of the recording head over the media. The latter has been reduced from 10s of μm in the first HDDs to 1-2 nm in current products, approaching atomic scales [2]. Finally, the number of magnetic grains per bit has been reduced from 1000s early on to about 20 in current products, approaching the practical Signal-to-Noise-Ratio (SNR) limits of signal retrieval. Further reduction in the number of grains can be achieved by using multiple readers and advanced signal detection schemes, an approach called two-dimensional magnetic recording (TDMR) [3].

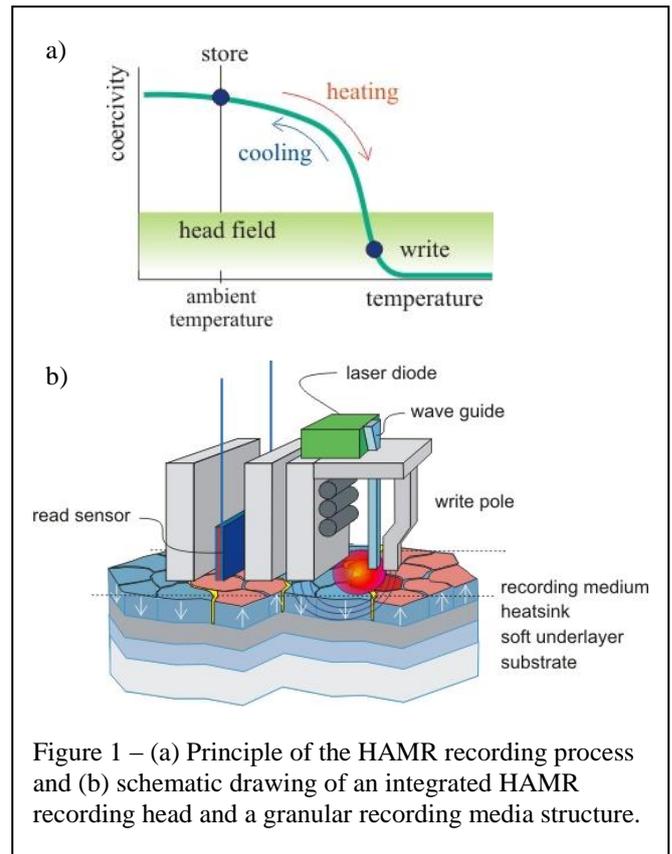

Figure 1 – (a) Principle of the HAMR recording process and (b) schematic drawing of an integrated HAMR recording head and a granular recording media structure.

Limits of the current recording technology are widely anticipated to be reached at about 1 Tb/in$^2$ or slightly above. One approach to push AD significantly beyond 1 Tb/in$^2$ is to reduce the number of grains to one lithographically defined magnetic dot, an approach known as bit patterned magnetic media recording (BPMR) [4]. While basic lithography capability up to 4 Tbit/in$^2$ has been demonstrated, BPMR requires enormous capital investment and has been pushed out on the HDD industry roadmap to 2020 or beyond.

**Advances in Science and Technology to Meet Challenges -** The approach widely considered the next HDD technology is called Heat Assisted Magnetic Recording (HAMR, also known as thermally or energy assisted magnetic recording). This recording scheme enables the use of granular media materials with very high $K_U$ that are thermally stable (but cannot be written by conventional recording heads) by locally heating the media and lowering its coercivity during writing, thus overcoming write field limitations, as shown in Fig. 1. A first HAMR recording demonstration at 250 Gb/in$^2$ was reported in 2009, and a recent recording demonstration of 1 Tbit/in$^2$ surpassed the highest AD achieved with conventional magnetic recording [5]. Fully functional HDDs using HAMR technology have been showcased by Seagate [6] and others. In addition to a number of engineering challenges such as incorporating or mounting a laser diode to the slider carrying the elements of the recording head (Fig. 1), and developing a robust tribological interface that can withstand the temperatures required for recording, the key fundamental challenges for HAMR are to develop (1) an efficient and robust optical near-field transducer (NFT, a device that concentrates light to spot sizes well below the optical diffraction limit), and (2) suitable low noise high-$K_U$ magnetic media.

(1) Initially, finding efficient NFTs provided the largest challenge to HAMR, but over the last 15 years groundbreaking work in near-field optics and plasmonics of nano-scale aperture and antenna structures has resulted in a number of competing designs that concentrate light with sufficient intensity into optical spots of <50 nm required for recording densities >1 Tbit/in$^2$ to heat the media to recording temperatures of 500°C or above [7].

(2) On the media side three main characteristics can be identified: (i) just as in conventional magnetic recording a small grain size and grain size distribution combined with thermal stability at the storage temperature are necessary to achieve high recording performance and long term data stability; (ii) the thermal properties of the media structure need to be tailored such that sharp thermal gradients can be achieved, e.g. by using efficient heat sink layers in the media (Fig. 1); (iii) the temperature at which the media can be written needs to be adjustable to a reliably accessible write temperature with a large gradient of the media switching field at that temperature.

The width of the transition between two bits in the media is then defined by the convolution of the temperature profile, the change in media switching field with temperature, the isothermal media switching field distribution, and the write field profile [8]. To achieve the maximum field gradient a tight overlap of the magnetic and optical profiles is required to suppress thermally induced self-erasure during the writing process, adding a 4$^{th}$ criterion to the recording trilemma described above [9].

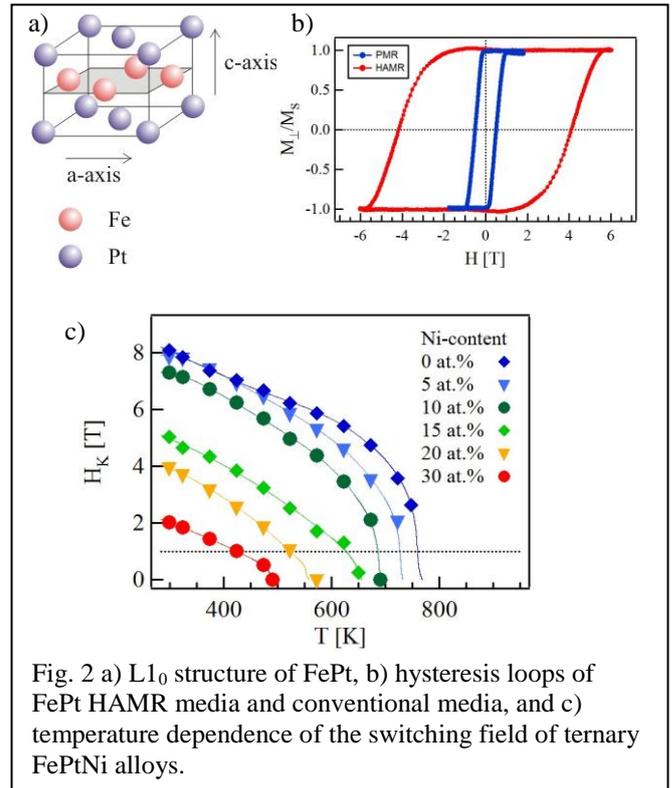

Fig. 2 a) L1$_0$ structure of FePt, b) hysteresis loops of FePt HAMR media and conventional media, and c) temperature dependence of the switching field of ternary FePtNi alloys.

The most promising high-$K_U$ materials for HAMR media are so-called L1$_0$ alloys based on chemically ordered FePt used in all recent HAMR recording demonstrations. Based on the $K_U$ of the fully ordered alloy grain sizes down to 3-4nm and AD up to 5 Tbit/in$^2$ seem feasible, however controlling grain size and grain size distribution at these sizes in a high-temperature deposition process required to achieve the chemically ordered L1$_0$ structure remain formidable challenges. Furthermore, while the high $K_U$ of FePt provides large thermal stability and potential AD gain, to sufficiently reduce the coercivity to enable writing requires heating the media close to their Curie temperature, $T_C$. For FePt $T_C$ is around 500 °C, posing significant challenges for the tribological head-disk interface. A lot of research is therefore being dedicated to materials such as ternary FeNiPt or FeCuPt L1$_0$ alloys shown in Fig. 2 [8], and to more complex layer structures [10], providing sufficiently high $K_U$ at reduced $T_C$ and recording temperature.

**Concluding Remarks** – With recent spin stand and drive technology demonstrations HAMR is well positioned for a full product introduction, with current predictions by industry experts ranging from 2016 to 2020 at AD >1Tb/in$^2$. Based on fundamental magnetic properties of the materials for the recording layer and currently known near field optical transducers an extension to several Tbit/in$^2$ seems feasible, beyond that a combination of HAMR and BPM is being considered to push magnetic storage technologies to its ultimate limits of 10 Tbit/in$^2$ or beyond [11].

# Developing synergies between organic and spin electronics

*Martin Bowen*, Institut de Physique et Chimie des Matériaux de Strasbourg, CNRS, University of Strasbourg

**Status** – The intersection between the fields of organic and spin electronics historically reflects a synergetic research compact. Indeed, spin-polarized currents are predicted to alter the singlet-to-triplet recombination ratio, and thus enhance the efficiency of organic light-emitting devices (OLED). Conversely, weak spin-orbit coupling in organic semiconductors (OS) is expected to promote long spin diffusion lengths and spin coherence times toward entire 'spintronics circuits'.

Initial measurements a decade ago [1] of spin-polarized diffusive transport across ~100nm-thick OS films underscore both the promise but also the immaturity of the field as controversy persists as to how this occurs [2]. Successful spin-polarized tunneling across OS barriers has, starting in 2007, demonstrated how to address the resistivity mismatch challenge toward spin injection into Oss [3]. Interfaces between ferromagnets and molecules with high spin-polarization beyond room temperature augur an imminent golden age for the field as it proposes the first candidate [4] for an ideal spin-polarized current source (see Fig. 1) after a 25-year search within spintronics at large. An important expected milestone was the demonstration of control over electroluminescence using spin-polarized currents (see Fig. 2) [5].

If the field can establish a solid knowledgebase of spintronics using simple molecules, then the wide array of molecular classes with intrinsic properties can be deployed to promote additional functionalities to spintronic devices. For instance, organic magnetic semiconductors may replace conventional electrodes [7], perhaps also barriers, in spintronic devices. Also, spin crossover molecules may promote memristance within a single-molecule junction [6], within within a more general vein of research on (spin-polarized) transport across single molecules.

**Current and Future Challenges** - Understanding the original observation of MR across thin (~100nm) OS films requires answering two questions:

*1) How are spins injected into the OS?*

Spinterfaces reveal a very rich feature set such as modifications to the underlying ferromagnet's anisotropy [8]. Furthermore, magnetic interactions within OS thin films such as metal phthalocyanines endowed with magnetic 1D chains of 3d spins could in principle interact with the spinterface property. Disentangling the impact of these effects on the effectiveness of an interface to promote spin injection will be crucial. Future work can exploit the unique magnetic anisotropy of the spinterface to tune the degree of spin polarization of the current flowing from/to spinterfaces. This notably requires a more systematic appraisal, for a given ferromagnet/molecule pair, of electronic properties using spectroscopic techniques and transport properties within actual devices. Can large values of tunneling magnetoresistance (presently 300-500% at T=2K [9]) be measured at room temperature per the spinterface's promising properties [4]? Can intrinsic molecular properties such as spin crossover [6] survive at a spinterface for enhanced functionality?

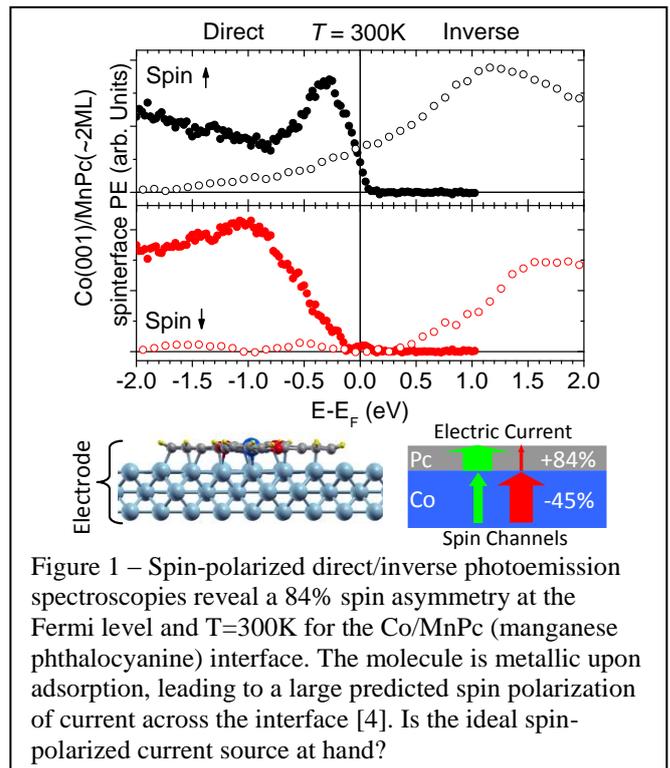

Figure 1 – Spin-polarized direct/inverse photoemission spectroscopies reveal a 84% spin asymmetry at the Fermi level and T=300K for the Co/MnPc (manganese phthalocyanine) interface. The molecule is metallic upon adsorption, leading to a large predicted spin polarization of current across the interface [4]. Is the ideal spin-polarized current source at hand?

*2) How do spins remain coherent over long diffusion lengths/times within OS layers?*

Organic magnetoresistance (OMR), the change in current flowing across a thin (~100nm) OS layer due to an external magnetic field, has been explained through numerous theories involving electron-hole pair interactions, triplet excitons, polaron pairs, bipolarons or even spin-orbit coupling within a magnetic field. Yet present data does not readily confirm one model over another. As proposed elsewhere [2], the challenge here is to undertake fundamentally sound scientific methodology that critically assesses the ensemble of results when promoting a given theory, with support from experiments whose physical underpinnings are solid, rather than the OMR effect itself. Notably, values of spin relaxation times as large as 1s have been reported, only afterwards to be attributed to artefacts such as the stray magnetic fields from the electrodes. This confirms the need for renewed methodological scrutiny.

**Advances in Science and Technology to Meet Challenges** - A recurring scientific difficulty when addressing the field from the (driving!) perspective of devices is how to account, in experimental results and their interpretation, for the structural quality of the organic layers and of the interfaces they form with the metallic electrodes. Indeed, substantial interdiffusion occurs when depositing a metal layer atop an organic layer. This is partly circumvented in organic electronics through 'dense layers' such as LiF or $MgF_2$. Avoiding interdiffusion could help better understand the OMR effect, though perhaps at the expense of leveraging spinterface properties.

Structural disorder in OS thin films leads to hopping transport across localized states, which complicates an understanding of diffusive spin transport. Beyond basic studies on bulk crystals, structural improvements in films and interfaces could clarify results in the field. Inherent to this point is the need to define what constitutes a reasonably ordered OS film.

To harness spinterface properties in devices, must one deposit the counterelectrode at low temperature [3]? Can macrojunctions synthesized using shadow masks [1] perform as well as do nanojunctions [9]? Imprint techniques may enable the synthesis of device stacks with bottom and top spinterfaces of pristine quality. Can existing devices processing technologies be adapted to solvent-averse OS? A mature ab-initio theoretical foundation that correctly describes both metals and molecules can now correctly predict the molecule's distance when adsorbed onto a metallic suface, and ensuing electronic properties [10]. Thus, the spinterface properties of ferromagnet/molecule pairs may be systematically calculated. In turn, those pairs with interesting properties may be experimentally studied so as to promote significant conceptual advances for the field and enable serious consideration from the industry.

Finally, many devices in the field rely on a ferromagnetic oxide (overwhelmingly $La_{0.7}Sr_{0.3}MnO_3$, or LSMO, see Fig. 2 and refs. [1,5,7,9]) as the lower electrode due to deposition conditions used for the organic layers (such as spin-coating in a glove-box atmosphere) that would otherwise oxidize a transition metal ferromagnet and thus reduce performance. Yet such electrodes aren't amenable to industrial applications [4], in particular because they do not work at room temperature. This in turn hobbles a fundamental understanding of spin-polarized transport at room temperature. The field should therefore focus on simpler spin injectors such as 3d transition metals, which anyways convey potential spinterface properties [4,8]. This implies a search for strategies to prevent/mitigate oxidation. In this respect, the ability for graphene to impede the oxidation of a ferromagnetic surface could prove crucial to the integration of effective spinterfaces within industrial processes.

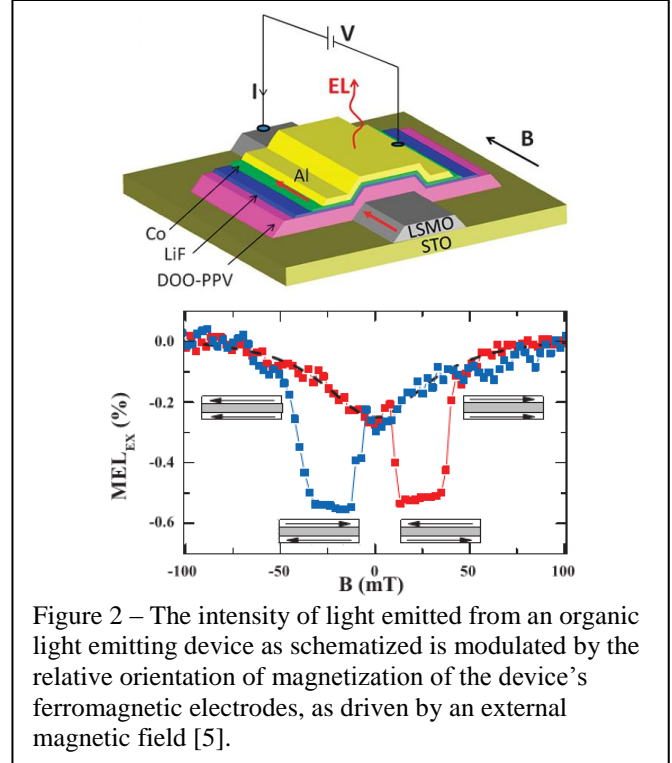

Figure 2 – The intensity of light emitted from an organic light emitting device as schematized is modulated by the relative orientation of magnetization of the device's ferromagnetic electrodes, as driven by an external magnetic field [5].

**Concluding Remarks** – The field of organic spintronics rose rapidly through spectacular initial experiments [1]. At present, certain milestones, both expected [5] and unexpected [4], have been attained. Controversy over our understanding of diffusive spin transport [2], and the deleterious effect on the perception of this research field, can be settled using a critical scientific methodology and established tools of physical investigation. The next Eldorado for the field, once it establishes a firm base of knowledge, will be to harness chemical engineering to design molecules with intrinsic properties (e.g. [6]) so as to further exalt the developing synergies between organic and spin electronics.

I acknowledge fruitful research with past and present collaborators, as well as funding from the ANR, CNRS, Institut Carnot MICA and U. de Strasbourg. Only a limited number of references could be included due to editorial constraints.

# Magnetic nanoparticles for biomedicine
*Sara A. Majetich*, Carnegie Mellon University

**Status** – Applications of magnetic nanoparticles in biomedicine rely on one of three key properties, as shown schematically in Figure 1. First, they can be moved by an inhomogeneous magnetic field as in magnetic separation, drug, or gene delivery. Second, they can be detected based on the local magnetic field they generate, as in magnetic resonance imaging, magnetic particle imaging, and magnetoresistive biosensing. Third, they can heat their local environment due to dissipation in an AC magnetic field magnetic fluid hyperthermia.

The nanoparticles used in vivo are almost exclusively made of superparamagnetic iron oxide, and coated so they are non-toxic, biocompatible, biodegradable, trigger no immune response, are not cleared too quickly, do not settle out of a dispersion, and do not form large aggregates [1], [2]. Particles or particle agglomerates < 100 nm are generally taken up by the reticuloendothelial system, while those > 5 μm can clog capillaries and trigger serious health problems. Nanoparticles used for ex vivo applications are generally micron-sized polymer composite beads containing a multitude of iron oxide nanoparticles. Unlike electric fields, magnetic fields have a minimal direct effect on most biological processes. Magnetic nanoparticles are used with external magnetic fields to target specific locations, to sense the local environment, and to controllably perturb that environment by delivering a stimulus such as heat, or a payload, such as a chemotherapy drug.

**Current and Future Challenges** – Magnetic separation is the most widely used application of nanoparticles in biotechnology [3]. Here magnetic beads are first functionalized to bind selectively to a toxin such as lead, or a food contaminant such as E. coli. After incubation in blood, high gradient magnetic fields are used to retrieve the particles. If done on a large scale, in analogy to kidney dialysis, it could be used for rapid decontamination after exposure to radionucleotides, chemical or biological weapons [4].

Magnetic separation is also being developed for lab-on-a-chip testing, in which a drop of blood is incubated with functionalized magnetic beads and then passed through a microfluidic device with regions surface functionalized to form sandwich assays [5]. For example, magnetoresistive sensors can be used to detect binding of particles bound to proteins associated with different kinds of cancer, in femtomolar concentrations [6]. The main challenges here are needs for improved specificity, dynamic range, and speed. Because magnetic separation can be done ex vivo, the particles are not restricted to be iron oxide. An intriguing new type of particle is a multilayer magnetic disk that acts as a synthetic antiferromagnet in the absence of an applied field, and therefore has no tendency to agglomerate [7]. As work continues in this area, the challenges are to reduce the size of the disk, functionalize them for in vivo use, and to develop methods to scale up production.

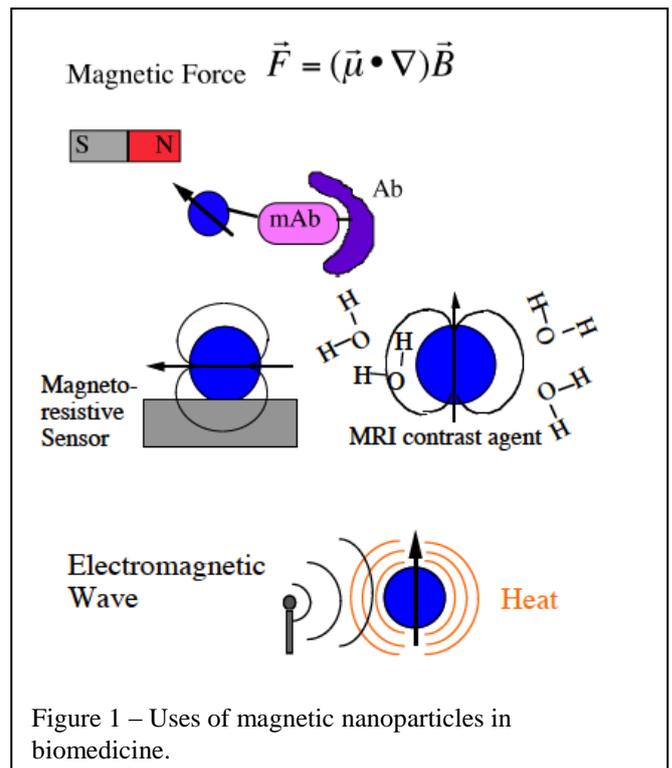

Figure 1 – Uses of magnetic nanoparticles in biomedicine.

Magnetic hyperthermia typically uses a 100 kHz - 1.2 MHz magnetic field to kill cancer cells close to magnetic nanoparticles due to local heating. The technique has recently been used for clinical trials to treat glioblastoma, and there have also been studies focused on head, neck, and prostate cancer [8]. Here the particles are injected directly, rather than relying on less invasive selective binding to cancer cells. Deep tumors present a challenge because of skin depth effects with AC electromagnetic fields. With a single source, the heating dose would be greatest at the surface, so methods to focus electromagnetic radiation are needed. Because the electromagnetic properties of muscle and fat are different, simulations of energy deposition are needed to avoid hot spots in fatty tissue.

Magnetic nanoparticles are playing an increasingly important role in biomedical imaging. The most common case involves small iron oxide nanoparticles as $T_2$-weighted contrast agents for magnetic resonance imaging (MRI), where their magnetic field causes the protons in water molecules to relax faster. Thalassemia, an iron overload disease, can be detected from MRI of the liver, without the need for painful biopsy [9]. Individual magnetic beads have been detected in microcapillaries, but greater sensitivity is needed for this spatial resolution within a person.

Depending on the applied magnetic field, the highest volume resolution is now a voxel 20-50 µm on a side. Magnetic particle imaging is a related technique using lock-in methods to detect the signal from individual magnetic bead dynamics, at ~50 frames/s [10], [11]. Alternative imaging strategies rely on direct detection of the field due to the particles. During breast cancer surgery functionalized particles are injected at the tumor site and the surgeon uses a gradient magnetometer to determine how far the cancer has spread into the surrounding lymph nodes [12]. While the spatial resolution is much lower than in MRI, it enables identification of the sentinel lymph node, which should be removed to minimize the chance of the cancer spreading.

**Advances in Science and Technology to Meet Challenges -** More efficient separation of circulating tumor cells (CTCs), a primary cause of cancer metastasis, would have an enormous impact on cancer treatment. Kaplan-Meyer curves show that the presence of >10 cancer cells/milliliter of blood correlates with a greatly reduced five-year survival rate [13]. Even with highly selective binding, this is a challenge, since the same milliliter contains ~5 x $10^9$ red blood cells and ~5 x $10^6$ white blood cells. Separation will require very high magnetic field gradients plus very high throughput.

The magnetic forces used for separation can also be used for in vivo magnetic targeting. Magnetic particles are now being investigated for stem cell delivery [14] as well as drug delivery that is triggered by the change in pH within cell endosomes [15].

More efficient selective targeting would benefit hyperthermia. Particles with a high degree of specificity for particular cell surface receptors often have a very low (~1%) targeting rate when introduced systemically. An alternative would be to improve methods for active magnetic guidance using multiple electromagnets [16] or tethered particles known as magnetic swimmers whose motion is controlled remotely [17]. Both of these advances would also benefit drug targeting, so that a chemotherapy agent could be given at a high local dose with fewer side effects. There are still questions about the mechanism by which hyperthermia kills cells, whether by heat generated by Brownian rotation or Néel relaxation of the particle, and on the response of biomolecules within the cell to local heating [18].

In magnetic imaging there have been demonstrations of microfabricated particles that have different resonant frequencies for MRI [19]. So far the tests have been in phantoms, but progress toward smaller, biocompatible particles suitable for in vivo use would enable more precise magnetic imaging.

While magnetic nanoparticles are now being used or investigated in both diagnosis and treatment of disease, there is emerging work using the particles to control the biological functions of cells or organisms. Bone cells (osteoblasts) grow faster when exposed to magnetic fields at low frequencies, and gene delivery to cell nuclei shows greater transduction efficiency [20]. The motion of C. elegans can be controlled with an AC magnetic field through local heating [21].

**Concluding Remarks –**

The fundamental magnetic properties of these monodomain particles --- forces, fields, and energy dissipation --- are well understood. Because of this they can be used as a tool for diagnosis, treatment, and addressing fundamental questions in biomedicine.

# Domain wall based devices

*Mathias Kläui*, Institute of Physics and Graduate School of Excellence Materials Science in Mainz, Johannes Gutenberg Universität Mainz, Mainz Germany

**Status:** In the past, magnetic nanostructures have been at the heart of a multitude of devices ranging from sensing applications to logic and data storage. The last comprises probably the best known magnetic devices that are the disc drive and magnetic tape. Both are based on the mechanical motion of the media that poses reliability and power consumption challenges and so a paradigm shift away from hard drives and tape is happening to novel solid state magnetic storage devices, such as the racetrack memory [1]. Furthermore such devices based on domain walls as discussed below are fast and non-volatile and could therefore replace all the different memory types used at the moment, thus radically simplifying the device architecture and are therefore being investigated in in academia as well as industry [1].

Conventional logic devices based on semiconductor gates are well established, however they are volatile and thus suffer from constant power consumption. Magnetic logic using movable magnetic domain walls as information has been proposed and the basic logical gates "AND" and "NOT" have been successfully demonstrated in magnetic nanowires [2].

The probably least known but (to our knowledge) so far only application based on domain walls that has made it to the market are magnetic sensors (see for example the RSM2800 at www.novotechnik.de) [3]. Domain wall based multiturn sensors are attractive for automotive and industrial application and can successfully compete with existing mechanical turn or angle sensors. Spiral sensors, such as the RSM2800 with limited turns are commercially available and may be used for instance to count, without mechanical wear, the turns of a steering wheel. A future design of a closed loop sensor has a huge potential due to the high number of turns that can be counted, which together with a suitable leverage could result in cheap and high precision angle sensors [3].

In all these applications, the size of the domain walls governs the achievable miniaturization and thus device density while the speed at which the domain wall can be manipulated governs the device performance. So in the next sections we will look at these two key parameters and analyze necessary developments to make the devices competitive.

**Current and Future Challenges:** For the operation of a memory, logic or sensing device, firstly well-defined domain walls are necessary and secondly the controlled manipulation of the domain walls is required.

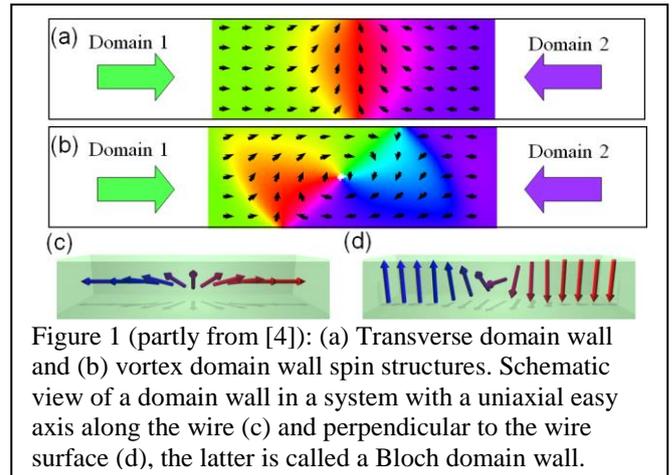

Figure 1 (partly from [4]): (a) Transverse domain wall and (b) vortex domain wall spin structures. Schematic view of a domain wall in a system with a uniaxial easy axis along the wire (c) and perpendicular to the wire surface (d), the latter is called a Bloch domain wall.

To use domain walls, the control and understanding of the domain wall spin structures is needed, which is most easily obtained in confined geometries such as nanowires where the spin structure can be tailored by the geometry as well as the material [4]. In general, the spin structure and thus the domain wall width is the result of an energy minimization process [4]. In the simplest case, i.e. without any externally applied fields and for materials without effective anisotropy (e.g. polycrystalline materials), the two important energy terms are the exchange energy and the stray field energy. In Fig. 1, we show typical examples of domain wall spin structures in such soft magnetic nanowires. Depending on the geometry one finds transverse (Fig. 1 (a)) and vortex walls (Fig. 1 (b)).

Additional contributions to the overall anisotropy energy can arise e.g. from interfaces (for instance in perpendicularly magnetized 3d metal / transition metal multilayers) or from an intrinsic magnetocrystalline anisotropy in epitaxially grown materials. In Fig. 1 (c) and (d) we compare schematically the spin structure for an easy axis along the wire (c) (as induced for instance by shape anisotropy as the case also in Figs. 1 (a) and (b)) and an easy axis perpendicular to the wire (d), where a Bloch wall spin structure is shown (for more details on domain wall spin structures, see [4-6]).

A key parameter is the domain wall width, which is determined by material parameters (exchange constant, saturation magnetization and anisotropies) as well as the geometry (that influences the stray field energy) [4-6]. It can range from hundreds of nanometers in soft magnetic materials down to a few nanometers in high anisotropy materials thus promising for applications.

To manipulate domain walls, different approaches have been put forward. For sensing and logic, conventional magnetic fields are used [6] and they have been shown to displace domain walls reliably [7] and quickly [8]. Beyond simple displacement, the domain wall spin structure can transform leading to oscillatory domain wall velocities in curved geometries below and above the Walker breakdown [7]. Recently a novel scheme has been put forward to move multiple domain walls with the same chirality synchronously by using perpendicular field pulses [9].

Due to more advantageous scaling, current-induced wall propagation has been most intensively investigated [1] (for a theoretical review see [10], and for an experimental review see [5]). The spin – polarized current exerts two torques: (i) the adiabatic spin torque where in the simplest approximation the current transfers one ℏ per electron across a 180° domain wall and thus the wall velocity scales with the spin polarization divided by the saturation magnetization for a given current density; (ii) the non-adiabatic spin torque that comprises different origins [5,10], and is governed phenomenologically by the nonadiabaticity parameter β and the velocity scales with β divided by the damping constant α (times the adiabatic torque strength). Recently further spin torques due to spin orbit coupling effects have been identified in multilayer materials [11-13], and these torques hold great promise for future applications as discussed in the next section.

**Advances in Science and Technology to Meet Challenges**: From the previous deliberations, it is clear that one needs a multi-pronged approach in order to further develop domain wall – based devices.

1. DW width and spin structure control:

In order to obtain devices with reliable operation, one needs reproducible DW spin structures and DW positioning.

To this end, one needs to optimize materials and nanofabrication to obtain domain walls where the spin structure is governed by the geometry and intrinsic materials properties rather than natural extrinsic pinning sites due to materials defects or roughness resulting from the nanostructuring process. A reduction of the natural pinning will lead to smaller excitations being able to manipulate the domain wall thus reducing the necessary fields and currents for logic and memory applications and increasing the operating window for sensors and allows for better defined artificial pinning sites for domain wall positioning.

However beyond optimizing currently available materials and their patterning onto the nanoscale, we need to develop new materials, which combine advantageous properties, such as high spin polarization and low saturation magnetization with well-defined domain walls. Examples that are promising include half metals in new materials classes, such as intermetallic Heusler compounds or oxides where recently well-defined spin structures were observed [5]. Additionally materials beyond ferromagnets, such as ferrimagnets and antiferromagnets hold large potential.

Furthermore as the domain wall width not only governs the minimum spacing between domains and thus for instance the storage density but is also a key parameter for the interaction between domain walls and fields [6] as well as spin-polarized currents [5,10], the wall width needs to be tailored by developing materials with the appropriate exchange interaction and anisotropies.

Finally further energy terms that allow one to engineer the domain wall spin structure can be invoked, such as the Dzyaloshinskii Moriya interaction that was shown to stabilize chiral spin structures [12] and set chiralities as for instance necessary for synchronous field-induced wall motion of multiple domain walls [9,11-13].

2. Develop new manipulation methods beyond fields and spin transfer torques:

While spin transfer torques entail good scaling for devices, the adiabatic spin torque leads to a maximum angular momentum transfer of about ℏ per electron and it is unclear how large the non-adiabaticity parameter β can be tuned. Recently new spin orbit torques have been observed [11-13]. While the origin is not completely clear, explanations based on effects, such as the spin Hall effect [12-13] and the Rashba Edelstein effect [11] have been put forward and these allow for potentially much more angular momentum transfer per charge current unit. If these torques can be understood, controlled and further enhanced, they can be a formidable means to obtaining current-induced domain wall motion overcoming the current speed limitations due to the sustainable charge current densities.

Related approaches use pure diffusive spin currents with no net charge currents and such pure spin currents can be generated by spin injection, where the spin current generation and the DW motion due to the spin current absorption can be spatially decoupled [14]. A very exciting source of spin currents is the spin Seebeck effect, where thermal spin currents can be used to manipulate DWs. Finally on the ultra-fast timescale, superdiffusive spin currents can be generated by fs excitations and manipulate DWs on the fs timescale [15] and further ideas based on magnonic spin currents and related approaches that warrant investigation have also been put forward.

Finally one needs to start thinking of new device concepts beyond the conventional memory, logic or sensing architectures. A simple example is the combination of multi-turn sensing by DW motion and utilizing the non-volatile DW storage of the information in the system that make additional semiconductor memory redundant.

**Concluding Remarks**: From sensors in the market to ideas for disruptive future applications, domain walls hold great promise for next generation devices. In the future, new materials with domain wall spin structures that can be tailored combined with new methods to manipulate them fast and efficiently will allow for the implementation of such devices. In particular with new spin orbit torques and pure spin currents, energy efficient manipulation with low charge currents becomes possible. With smart geometries and device architectures that for instance minimize DW interaction by introducing synthetic antiferromagnets one can reach highly integrated circuits that will be competitive with incumbent approaches.

Finally I would like to point out that there are hundreds of groups working on domain walls and due to the limited space, only a very small selection of mostly review references could be included, which are meant to serve as a source for further reading. I gratefully acknowledge the support of the past and present group members and in particular M. Foerster, T. Moore, O. Boulle and R. Mattheis as well as generous funding by the DFG and the EU.

# MRAM: status and roadmap
*I.L.Prejbeanu, B.Dieny*, Spintec

**Status of emerging nonvolatile MRAM market**

Demand for on-chip memories has been recently increasing due the growth in demand for data storage and the increasing gap between processor and off-chip memory speeds. One of the best solutions to limit power consumption and to fill the memory gap is the modification of the memory hierarchy by the integration of non-volatility at different levels (storage class memories, DRAM main working memory, SRAM cache memory), which would minimize static power as well as paving the way towards normally-off / instant-on computing (logic-in-memory architectures) (Fig.1). Besides computers, today's portable electronics have become intensively computational devices as the user interface has migrated to a fully multimedia experience. To provide the performance required for these applications, the actual portable electronics designer uses multiple types of memories: a medium-speed random access memory for continuously changing data, a high-speed memory for caching instructions to the CPU and a slower, nonvolatile memory (NVM) for long-term information storage when the power is removed. Combining all of these memory types into a single memory has been a long-standing goal of the semiconductor industry, as computing devices would become much simpler and smaller, more reliable, faster and less energy consuming. As a result, advanced NVM chips are expected to see phenomenal growth in the forthcoming years. MRAM is one of a number of new technologies aiming to become a "universal" memory device applicable to a wide variety of functions. MRAM have however not yet reached large volume applications, with only Toggle switching-based standalone products currently available from Everspin (Fig.2a) [1]. The more recent advent of spin transfer torque (STT) [2], however, has shed a new light on MRAM with the promises of much improved performances and greater scalability to very advanced technology node. Indeed, in 2010, the International Technology Roadmap for Semiconductors (ITRS), Emerging Research Devices and Emerging Research Materials Working Groups identified STTRAM as one of the two emerging memory technologies (with RRAM) recommended for scaling of non-volatile RAM to and beyond the 16nm generation. Start-ups, large IC manufacturers and equipment suppliers are now actively developing the STTRAM technology and a forthcoming launching of 64Mbit in-plane magnetized STTRAM products was recently announced by Everspin.

**Current and Future Challenges for MRAMs**

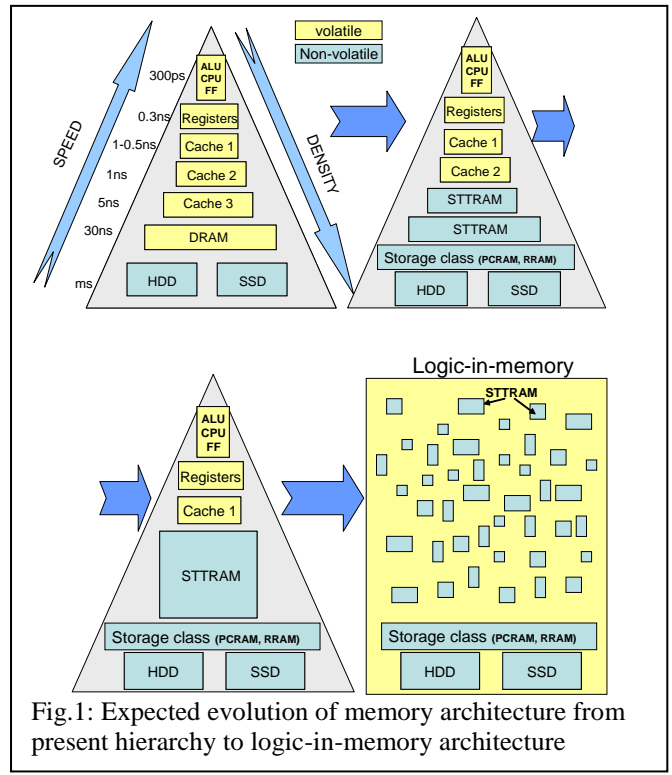

Fig.1: Expected evolution of memory architecture from present hierarchy to logic-in-memory architecture

MRAM technologies (Fig.2) evolved in the last years, benefiting from the progress in spintronics research, namely the tunnel magnetoresistance (TMR) of MgO magnetic tunnel junctions (MTJ) [3], the STT [2] and the spin orbit torque (SOT) [4] phenomena. The elementary cell of all these MRAM architectures is a MTJ consisting of two ferromagnetic layers separated by a thin insulating barrier and the readout (i.e. determining the magnetic sate of the MTJ) is always performed by measuring the MTJ resistance. An extension of the initial field written MRAM (Fig.2a) is the Thermally Assisted MRAM (TA-MRAM) [5] (Fig.2b), wherein the write selectivity is achieved by a combination of a temporary heating produced by the tunneling current flowing through the cell and of a magnetic field. Besides, TA-MRAM with a soft reference allows introducing new functionalities such as the "Match In Place™", particularly promising for security and routers applications. However, the downsize scalability in conventional field-writing technology is limited to about 60nm, due to electromigration issues in the field lines. STT writing (Fig.2c) offers a much better downsize scalability as the critical current for writing decreases proportionally to the cell area down to a minimum value set by the retention (~15µA). The greatest interest is now focused on out-of-plane magnetized STTRAM (p-STTRAM), taking advantage of the perpendicular magnetic anisotropy which exists at the CoFeB/MgO interface (Fig.2d) [6]. p-STTRAM require significantly less write current than their in-plane counterparts for a given value of memory retention and provide a better stability of the written information. Optimized p-STTRAM stacks will likely comprise two tunnel barriers with antiparallel polarizing layers to maximize anisotropy and STT efficiency (Fig.2e). The thermal assistance can also be combined with STT to circumvent a classical dilemma in data storage between the memory writability and its retention [7]. Recently it has been shown that assistance by an electric field may reduce the STT writing critical currents in MTJs [8]. This has been demonstrated in magnetic stacks with perpendicular magnetic anisotropy but the effect is quite weak when using metallic layers due to the electric field screening over the very short Fermi length in metals. An electrically reduced magnetic anisotropy leads to lower energy barrier that is easier to overcome for changing magnetization direction. In principle, voltage control spintronic devices could have much lower power consumption than their current-controlled counterparts provided they can operate at sufficiently low voltage (typically below 1V). Multiferroïc or ferromagnetic semiconductor materials could provide more efficient voltage controlled magnetic properties. 3-terminal MRAM cells written by domain wall propagation (Fig.2f) or SOT (Fig.2g) were also recently proposed [4,9] to separate write and read current paths. This can ease the design of non-volatile logic circuits and increase the reliability of the memory. SOT-MRAM offers the same non-volatility and compliance with technological nodes below 22nm, with the addition of lower power consumption (provided the write current density can be further reduced thanks to stack optimization), cache-compatible high speed and improved endurance. The drawback is the increased cell size.

**Advances in Science and Technology to Meet Challenges**

MRAM technologies are still in their nascent stages, particularly at the sub-20nm dimensions. There is still a need to strengthen the technology maturity and for advances in circuit designs and innovative architectures. The main issues remain associated with the cell to cell variability, TMR amplitude and temperature range. Variability is mainly caused by edge defects generated during patterning of the cells. MgO damages yield local changes in the barrier resistance, TMR and magnetic anisotropy i.e. cell retention. With the increasing number of actors now working on this technology, faster technological progresses can be expected in the near future. Also, implementing self-referenced reading scheme can lead to improved tolerance to process defects. Concerning out-of-plane STTRAM, progresses are needed in the composition of the stack to minimize the write current, maximize the TMR amplitude and improve the temperature operating range. Double barrier MTJ with separately optimized interfacial and bulk properties

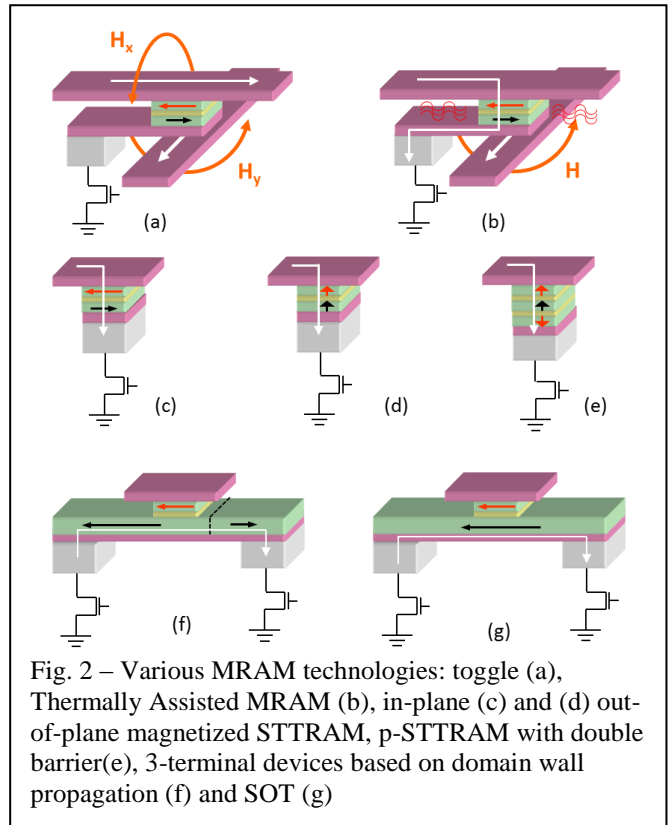

Fig. 2 – Various MRAM technologies: toggle (a), Thermally Assisted MRAM (b), in-plane (c) and (d) out-of-plane magnetized STTRAM, p-STTRAM with double barrier(e), 3-terminal devices based on domain wall propagation (f) and SOT (g)

should allow reaching the requirements for Gb STTRAM at 12nm node (Fig.2e). Heusler and $X_{1-x}Mn_x$ (X = Cr, V, Ge, Ga…) alloys have also already demonstrated their potential for p-STTRAM (low $M_s$, large perpendicular anisotropy, low damping) [10] but none of the existing alloys combine all required properties yet.

**Concluding Remarks**

MRAM is viewed as a credible replacement to existing technologies for applications where the combination of non-volatility, speed and endurance is key. This technology is receiving an increasing interest in the microelectronics industry. In particular, STTRAM has the potential of delivering high density and a scalable technology down to size ~10nm by using out-of-plane magnetized MTJ. The ultimate scalability in STTRAM could be provided by combining thermally/voltage assisted switching and STT. SOT-MRAM can be viewed as a very interesting approach for non-volatile logic and MRAM of improved endurance. Voltage controlled spintronics devices may later yield devices of much reduced power consumption. As a matter of fact, there is lot of room for reducing the power consumption in MRAM technologies considering that the barrier height to insure a 10 year retention of a Gb chip is typically of $80k_BT \sim 4 \times 10^{-4}$fJ whereas the energy presently required per STT write event is in the range 50fJ-1pJ.

# Permanent magnets

*Nora M. Dempsey*, Institut Néel – CNRS/UJF, Grenoble

**Status –** The development of permanent magnets over the last century can be gauged by the growth in the value of maximum energy product, $(BH)_{max}$, a figure of merit that corresponds to twice the work that can be done by a magnet (figure 1). Successive increases were due to the discovery of new materials with improved intrinsic magnetic properties and the development of appropriate microstructures through new and optimized processing techniques [1]. The $(BH)_{max}$ achieved in alnico magnets are limited by their low values of coercivity, determined by shape anisotropy. Exploitation of magneto-crystalline anisotropy in ferrite magnets led to higher values of coercivity, but their magnetization and $(BH)_{max}$ are limited by their ferrimagnetic nature. The most significant breakthrough was the discovery and development of rare earth - transition metal (RE-TM) magnets, which owe their high values of coercivity to very high values of magneto-crystalline anisotropy.

The market share of the different magnets, determined by the balance between cost and performance, is shown in the inset of figure 1. While steel magnets are now obsolete, alnicos are still used where the temperature stability of magnetization is important. Cheap ferrite magnets account for the greatest tonnage of magnets produced annually, being used where their magnetic properties are sufficient (e.g. in small electrical machines, latches…). RE-TM magnets have revolutionized the design of motors, generators and actuators. Sm-Co magnets are used where very high coercivities or very high temperature tolerances (> 200°C) are required. Nd-Fe-B magnets are used where the volume and or weight of the flux source must be minimized. Firstly mass-produced for use in the voice coil motor of hard disk drives, they are now a key component of the electric motor and generator of (hybrid) electric cars and the generator of direct-drive wind turbines. For the latter applications, heavy rare-earths such as Dy partially replace Nd to maintain sufficient coercivity at the elevated operating temperatures (<180°C). Finally, other factors such as mechanical, electrical or corrosion characteristics may also determine the choice of magnet type for a particular application.

**Current and Future Challenges -** While a major achievement would be to find new hard magnetic materials with properties surpassing those of Nd-Fe-B, much can be done to improve the performance and reduce the cost of known materials. The coercivities achieved in magnets are typically a fraction of the intrinsic upper limit given by the anisotropy field, due to microstructural defects, especially at grain surfaces. In the first generation of high-coercivity Dy-containing Nd-Fe-B magnets for hybrid cars and turbines, Dy increases the magneto-crystalline anisotropy of the main phase, and thus the magnet's coercivity, but reduces it's magnetization due to antiparallel coupling with Fe. Moreover, Dy is rarer and typically ten times more expensive than Nd, and both price volatility and concerns over sourcing have led to a major drive to reduce or eliminate the need for it. This poses the challenge to reduce the difference between coercivity and anisotropy by combatting the initiation and propagation of magnetization reversal at grain surfaces, through grain boundary engineering. Diffusion of Dy to the outer shell of $Nd_2Fe_{14}B$ grains results in high coercivities with much reduced overall Dy content [2]. The coercivity of Nd-Fe-B magnets is also being increased through grain size reduction combined with enhanced magnetic decoupling of neighbouring grains with non-magnetic grain boundary phases (e.g. Nd-Cu) [3]. The challenge now is to reduce the thickness of the secondary phase (high-anisotropy shell, non-magnetic grain boundary phase) to an absolute minimum. Another focus of Nd-Fe-B-based research concerns RE-lean hard-soft nano-composites in which the high magnetisation soft phase serves to increase the magnet's overall magnetization [4]. The challenge here is to achieve sufficient coercivity in textured structures. There is also much potential for the development and use of low-cost magnets with energy products intermediate between those of ferrite and RE-magnets [5]. To this end, a range of non-cubic Mn, Fe and Co based alloys are being studied. Here the challenge is to produce materials with a high volume content of a stable high anisotropy phase. Bulk materials with reasonable values of coercivity and energy product are being produced using hexagonal MnBi, while the properties of tetragonal MnAl bulk powders are also being improved. Tetragonal FeNi is being studied, though the weak anisotropy of this phase and it's slow kinetics of formation (it has been found in trace quantities in meteorites) poses major questions about its use in magnets. The promotion of interstitial Fe-N phases as magnet candidates is also questionable, again based on their metastability and weak anisotropy values. More promising results have been achieved recently in Co-rich metastable phases (Co-Hf, Co-Zr) [6]. Electronic structure calculations predict high uniaxial anisotropy in tetragonally distorted Fe-Co-W alloys, and are now being used for the in-silico search for new RE-free permanent magnet materials based on ternary and quaternary phases.

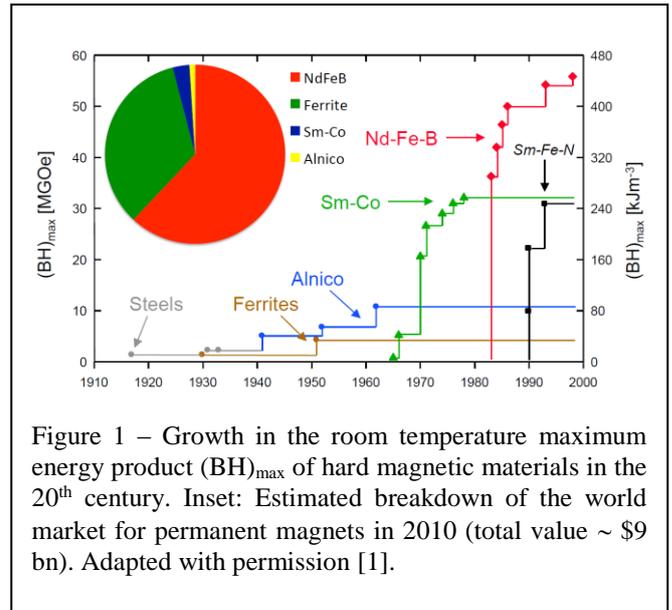

Figure 1 – Growth in the room temperature maximum energy product $(BH)_{max}$ of hard magnetic materials in the 20$^{th}$ century. Inset: Estimated breakdown of the world market for permanent magnets in 2010 (total value ~ $9 bn). Adapted with permission [1].

**Advances in Science and Technology to Meet Challenges** – Whether it is to improve the performance of known materials, or to develop new types of magnets, the essential challenge is to understand the link between coercivity and microstructure. This requires advances within, and synergy between processing, characterization and modelling (Figure 2). Recent developments in high-resolution structural characterization techniques, including 3D atom probe tomography, aberration-corrected transmission and scanning transmission electron microscopy combined with electron energy loss spectroscopy, are giving unprecedented access to the atomic structure and local chemical composition at surfaces and interfaces of magnets [3,7]. Developments are also being made in magnetic imaging of hard materials with techniques including scanning transmission x-ray microscopy for element-specific high resolution magnetic domain imaging, Lorentz TEM for magnetic domain wall imaging and magnetic induction mapping, and small angle neutron scattering for indirect analysis of magnetic domains in the bulk of the material.

Multi-scale modelling is being developed to simulate magnetization reversal in permanent magnets [8]. Ab initio electronic structure calculations of the intrinsic magnetic properties of the different magnetic phases present and molecular dynamics simulations of lattice distortions at grain interfaces are being fed into micromagnetic simulations of assemblies of hundreds of grains to predict the influence of defects and demagnetizing fields on bulk hysteresis properties. Spectacular progress is being made due to recent increases in available computing power through the use of massively parallel hardware such as graphic cards and improvements in numerical techniques based on finite element models.

A significant development in the fabrication of Nd-Fe-B magnets concerns the recently reported "pressless" process, in which oxygen pick-up by fine powders is minimized by carrying out all steps under a well controlled atmosphere [9]. Mechano-chemical and electro-chemical synthesis holds potential for the fabrication of nano-powders for use in magnets while the development of low temperature compaction processes (e.g. spark plasma sintering) is needed to minimize grain growth so as to preserve coercivity. Permanent magnet films now being developed have many potential applications in micro-systems. They are already being integrated into proto-type devices for bio-medical applications and in the future they could be exploited in micro-devices (sensors, actuators, energy harvesters…) that serve in energy management.

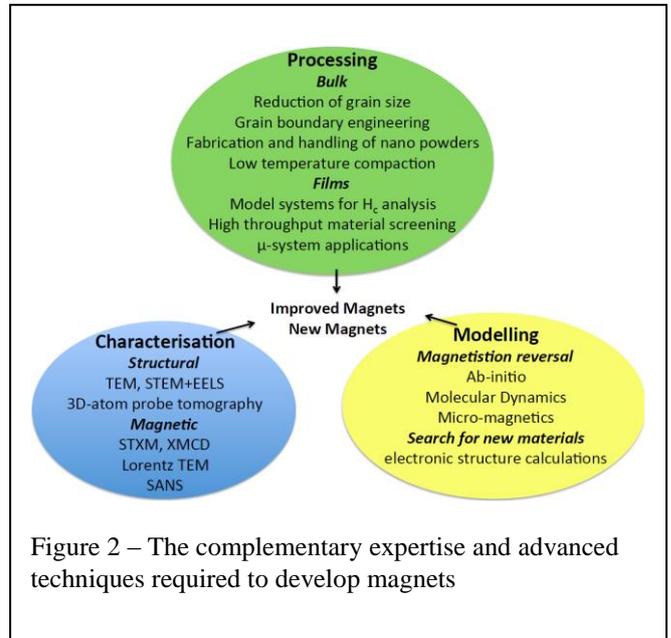

Figure 2 – The complementary expertise and advanced techniques required to develop magnets

**Concluding Remarks** – The recent RE-crisis has spurred a renewed interest in permanent magnet research. Significant developments in both characterization and modelling are proving very timely and are making a real contribution to improving our understanding of coercivity. The achievement of high values of coercivity in Dy-free Nd-Fe-B thick films [10], suggests that further progress will be made in reducing and eventually eliminating the need for heavy rare earths such as Dy in high-coercivity Nd-Fe-B-based bulk magnets. Progress in the fabrication of RE-TM nano-powders holds promise for the future development of high energy product hard-soft nano-composites. There is real scope for the development of new mid-range RE-free magnets, based either on nano-scaled composites exploiting shape and possibly surface anisotropy or on new phases, the study of which could be guided by data-mining computational studies.